\documentclass[twocolumn,usenatbib]{mnras}
\usepackage{amsmath}

\usepackage{graphicx}	
\usepackage{amsmath}	
\usepackage{amssymb}	
\usepackage{multicol}        
\usepackage{bm}		
\usepackage{pdflscape}	
\usepackage[T1]{fontenc}
\usepackage{ae,aecompl}
\usepackage{newtxtext,newtxmath}
\usepackage{caption}
\usepackage{multirow}
\usepackage[utf8]{inputenc}
\usepackage{hyperref}
\usepackage{xcolor}
\usepackage{multirow}

\usepackage[T1]{fontenc}
\usepackage[english]{babel}
\usepackage[flushleft]{threeparttable} 
\usepackage{booktabs,caption}
\usepackage{blindtext}
\usepackage{subcaption}
\usepackage[flushleft]{threeparttable} 
\usepackage{booktabs,caption}
\usepackage{natbib}

\begin{document}

\title{Application of time-series analysis methods to a multiple-sector TESS observations: the case of the radio-loud blazar 3C 371}
\author[Tripathi et al.]{
Ashutosh Tripathi,$^{1,2}$\thanks{E-mail: ashutosh31tripathi@yahoo.com}
Paul J. Wiita,$^{3}$
Ryne Dingler,$^{2}$
Krista Lynne Smith,$^{2}$
R.\ A.\ Phillipson,$^{4}$
\newauthor
Matthew J.\ Graham,$^{5}$
and Lang Cui,$^{1,6,7}$
\\
$^{1}$Xinjiang Astronomical Observatory, CAS, 150 Science-1 Street, Urumqi 830011, People's Republic of China\\
$^{2}$George P. and Cynthia Woods Mitchell Institute for Fundamental Physics and Astronomy, Texas A\&M University, College Station, TX 77843-4242, USA \\
$^{3}$Department of Physics, The College of New Jersey, 2000 Pennington Rd., Ewing, New Jersey 08628-0718, USA\\
$^{4}$Department of Physics, Villanova University, Villanova, PA 19085, USA\\
$^{5}$California Institute of Technology, 1200 E. California Blvd, Pasadena, CA 91125, USA\\
$^{6}$State Key Laboratory of Radio Astronomy and Technology, A20 Datun Road, Chaoyang District, Beijing, 100101, P. R. China\\
$^{7}$Xinjiang Key Laboratory of Radio Astrophysics, 150 Science 1-Street, Urumqi 830011, P.R.China\\
}


\pubyear{2025}
\label{firstpage}
\pagerange{\pageref{firstpage}--\pageref{lastpage}}
\maketitle
\begin{abstract}

We present various time series analysis methods to analyze multiple-sector observations of bright AGN from the Transiting Exoplanet Survey Satellite (TESS) and examine whether issues such as gaps and noise in these data can be mitigated. We determine variability timescales and search for quasi-periodicity using these methods and assess any differences. In this paper, we present an analysis of the $\approx$~ 300-day TESS observation of a blazar 3C 371 using power spectrum density, structure-function, and weighted wavelet Z-transform approaches. To reduce the effect of gaps and noise, Continuous auto-regressive moving averages, Bartlett periodogram, and wavelet decomposition methods are used. We have also used recurrence analysis to account for the nonlinearity present in the data and to quantify variability or periodicity as the recurrent state. Considering the entirety of the TESS observations, we derive the variability timescale to be around 4.5 days. Sector-wise analysis found variability timescales in the range of 3.0--7.0 days, values that are found to be consistent using different methods. When analyzing multiple sectors together, significant variability, which could be quasi-periodic oscillations (QPOs), of duration 3--6 days in individual segments, is detected. These may be attributed to the kink instabilities developed in the jet or the existence of mini-jets inside a jet undergoing precession. We find that these methods, when applied appropriately, can be used to study the variability in TESS data. The noise present in these TESS observations can be minimized using Bartlett's periodogram and wavelet decomposition to recover the real stochastic variability.
\end{abstract}


\section{Introduction} \label{sec:intro}

Supermassive black holes (SMBHs) have masses between $10^5$--$10^{10}$ M$_\odot$ and are believed to be found in the centers of nearly all galaxies.  When accretion onto SMBHs yields luminous central regions, sometimes even more powerful than the integrated light of galaxies' stars, these systems are known as Active Galactic Nuclei (AGNs). 
AGNs are normally classified according to their optical, radio, and X-ray properties and their inclination with respect to the observer \citep{1989AJ.....98.1195K, 1995PASP..107..803U}. Blazars are radio-loud AGNs with one of their jets pointed towards the Earth. They are further classified into two classes on the basis of emission lines. The spectra of flat-spectrum radio quasars (FSRQs) exhibit broad emission lines, whereas the spectra of BL Lac objects exhibit no or weak emission lines and almost all the emission comes from the jet\citep[see][and references therein]{2008Natur.452..966M, 2019ARA&A..57..467B}.

Blazars are known to display variability on timescales ranging from a few minutes and hours to weeks, months, and even a few years across the whole electromagnetic spectrum \citep[see][and references therein]{1995ARA&A..33..163W, 2002A&A...390..407V,2005A&A...436..799F, 2010MNRAS.404.1992R,2021ApJ...923....7B, 2021MNRAS.501.1100R}. Different timescales in different wavebands can correspond to different physical processes and emission mechanisms. Therefore, variability is a very important tool for understanding the physical processes occurring within the accretion disks and the relativistic jets of blazars. However, observations usually combine the footprints of other random processes along with the possible physical signal,  which becomes very difficult to disentangle and interpret. Hence, observations that focus on long-term, short-term,  as well as intraday variability (IDV) can allow us to probe different processes on different timescales and possibly differentiate them from random uncorrelated processes \citep[see][and references therein]{2013MNRAS.431.1914G, 2020ApJ...903..134C,2023ApJ...951...58W}.

The Transiting Exoplanets Survey Satellite \citep[TESS;][]{2015ESS.....350301R} is a space-based optical telescope that provides very high-cadence data and its observations of a given area extend over temporal baselines that range from 27 to 355 days. Such observations are very suitable for investigating timescales of the order of a few hours to days and sometimes up to a few months. Unlike ground-based instruments, which are affected by daily and annual gaps, TESS provides more uniformly sampled data for variability studies. Each TESS Cycle covers either the northern or southern hemisphere and consists of 13 sectors of $\approx$ 27 days each. 

\begin{figure*}
\hspace{-1.5cm}\includegraphics[scale=0.8]{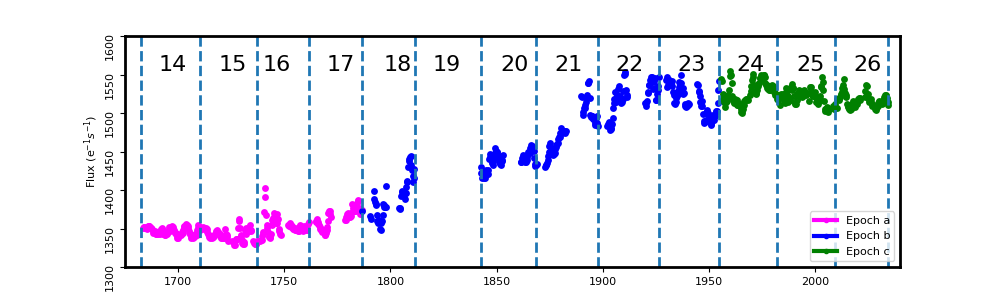}\\
\hspace{-1.5cm}\includegraphics[scale=0.8]{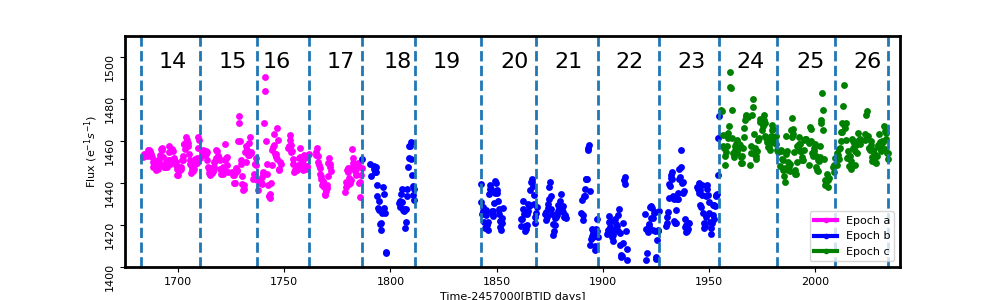}
\caption{Cycle 2 TESS observations of 3C 371 using simple (top) and fully (bottom) hybrid methods. The data is binned to 0.3 days for visual clarity. The entire Cycle 2 observation is divided into various sectors (by dashed vertical lines) whose numbers are indicated in the plot. In our analysis, we also divide the light curve into 3 epochs, A, B, and C, denoted by magenta, blue, and green, respectively.}
\end{figure*}\label{fig:lc}

The variability of stochastic processes, as observed in AGNs, is often described in terms of the power spectral density (PSD) and structure function (SF). The PSD measures the variability amplitude per temporal frequency, while the SF estimates the variability amplitude as a function of timescale. We also employ a recurrence plot analysis \citep[ RA;][]{2007PhR...438..237M}, which is a method that correlates positions in time to nearby positions in the phase space of a dynamical system, compiled into a square matrix. The resultant matrix can be visualized as a ``Recurrence Plot” (RP) with patterns unique to different dynamical systems. For example, repeating patterns in an RP is an indication of the presence of deterministic variability, possibly including quasi-periodicity, in the observation. The summation of an RP along the diagonal can be interpreted similarly to an auto-correlation function, which essentially summarizes the correlation between data points in one dimension for different timescales. RPs were originally utilized to study nonlinear heart rate variability and are now widely used in a diverse range of disciplines from physiology to geology. RPs fundamentally probe the higher-order, non-linear patterns of variable sources and enable one to distinguish a stochastic process from a deterministic process as responsible for the observed variability. 


The TESS data usually have gaps in the middle of each roughly monthly observation when it turns toward the Earth to transmit the data, as well as the gaps between adjacent sectors \citep{2023ApJ...958..188S}. To analyze multiple sectors of TESS observations, it is important to treat the gaps with care so that they do not significantly affect the measurements of any variability timescales. There are three approaches to account for the irregular sampling in the astronomical data. The first approach is to fit the light curve in the time domain and then estimate the evolution of the light curve by solving multiple-order differential equations. The solution of these differential equations predicts the observations in the gaps. To do this, we consider a continuous auto-regressive moving average (C-ARMA) process \citep[e.g.][]{2014ApJ...788...33K}. The second approach is to fit a periodogram to the power as a function of frequency using a red-noise model such as a simple or broken power law and compare the observed light curve with an ensemble of those having the same statistical and spectral properties using Monte Carlo simulations \citep{2013MNRAS.433..907E}. The third approach is to average the periodogram of the continuous part of the data, commonly known as the Bartlett periodogram \citep{1948Natur.161..686B}. 
In this work, we employ the wavelet decomposition (WD) method \citep{2001C} to reduce random noise and consequently increase the strength of signals at lower frequencies. We used these methods to account for gaps and random noise in the TESS light curves in a fashion that should be very useful in analyzing TESS data from other sources as well as from other future telescopes.

Here we analyze a $\approx$ 300d TESS observation of 3C 371, which is an object that has been considered both a radio galaxy and a BL Lac \citep{1975ApJ...200L..55M} and has a redshift of 0.05 \citep{1992A&AS...96..389D}. This object displayed remarkable variability across the whole electromagnetic spectrum on various timescales. It showed long-term variability during EXOSAT X-ray observations \citep{1990ApJ...356..432G} and timescales of several years in the ultra-violet \citep{1994A&A...291...74P}, $\gamma$-ray \citep{2020ApJS..247...33A}, and radio \citep{2009AJ....137.5022N} bands. In optical bands, 3C 371 shows both intra-day variability (IDV) \citep{2006A&A...448..143X} and timescales of days \citep{1967ApJ...150L...5O}. Interestingly, it also showed radio variability of the order of 2d \citep{2003A&A...401..161K}, which is consistent with the 2.4d timescale found in an optical R-band observation \citep{1996A&A...305...42H}. 

In Section ~\ref{sec:obs}, we describe the observations and data reduction. In Section~\ref{sec:carma}, we discuss various data analysis methods used in this work. In Section~\ref{sec:res}, PSD, SF, and RA estimates are presented for each sector individually and also for various segments of the Cycle 2 observations. We also present variability timescales estimated using different methods and then discuss different techniques used to mitigate gaps. We also search for possible quasi-periodic oscillations (QPOs) present in these epochs. In Section~\ref{sec:dis}, we summarize our results and present our concluding remarks in Section~\ref{sec:con}. 


\begin{figure*}
\hspace{-1.0cm}\includegraphics[scale=0.4]{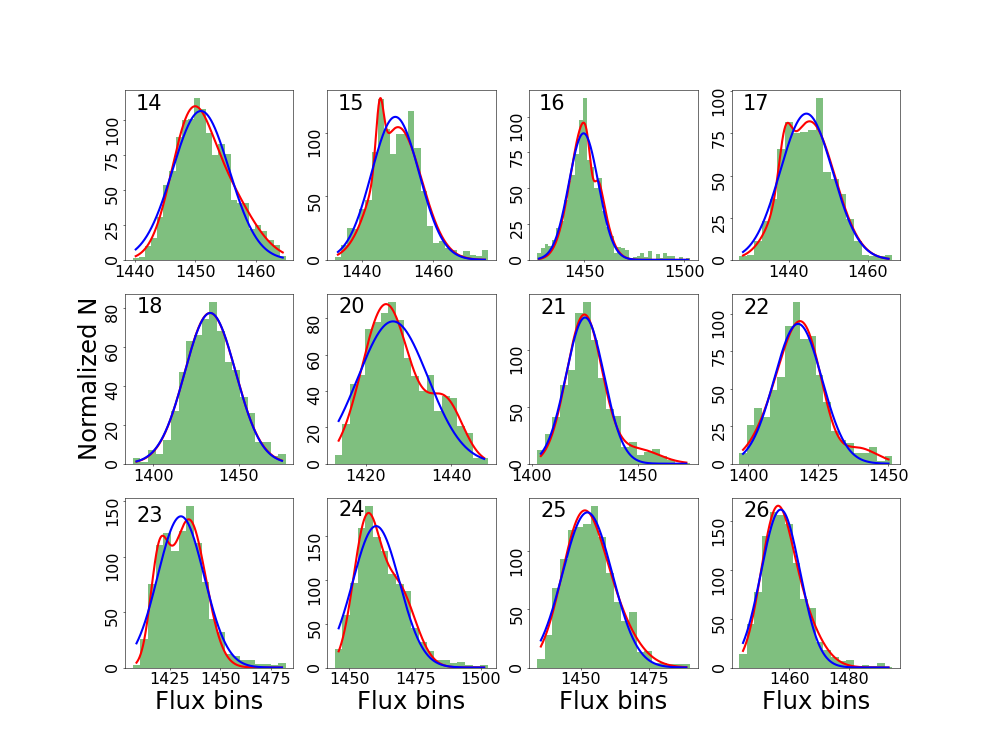}
\caption{Flux distribution of each sector of Cycle 2 TESS observations of 3C 371. Fits with Gaussian (blue) and bimodal Gaussian (red) distributions are also plotted.}
\end{figure*}\label{fig:flux}

\section{Observations and data reduction} \label{sec:obs}
3C 371 was observed by TESS between 19 July 2019 and 3 July 2020 corresponding to TESS Sectors 14 through 26, with the exception of Sector 19. TESS is primarily designed for searching for exoplanets which would impress periodic behavior onto stellar light curves via their transits \citep{2015ESS.....350301R}. As such, most TESS pipelines are not well-equipped to accommodate stochastic, aperiodic behavior.
Extracting AGN light curves requires careful treatment of the background sources and systematics, which could affect the light curve, apart from avoiding overfitting AGN variability with the periodic model used for exoplanets. Techniques for obtaining AGN light curves from very similar {\tt Kepler} data were considered in \citet{2018ApJ...857..141S}. We use the {\tt Quaver}\footnote{\url{https://github.com/kristalynnesmith/quaver}} pipeline \citep{2023ApJ...958..188S}, which was designed specifically to avoid over- or under-fitting of stochastic AGN light curves from TESS.
Using the TESSCut package \citep{2019ASPC..523..397B}, full-frame images (FFIs) for a particular observation can be used to extract the aperture of the desired object so as to exclude contamination from other sources. Once the desired aperture is selected, {\tt Quaver} implements several tasks provided by the Lightkurve \citep{lightkurve} package. These include a principal component analysis (PCA) to design matrices that account for the additive background flux and the multiplicative systematics affecting the source. Then, these matrices are used to correct the light curve of the source using Lightkurve's {\tt Regressioncorrector}. \citep[For more details, see][and references therein.]{2023ApJ...958..188S}

The {\tt Quaver} pipeline provides three distinct methods to correct for the systematic effects affecting the TESS light curves; these are simple PCA, simple hybrid, and full hybrid methods. The PCA method is the simplest and applies a user-specified number of principal components to correct the light curve. However, this method cannot correct for most of the instrumental systematics, and it also overfits the long-term variability present in the light curve. The simple hybrid method takes into account the background additive effects and instrumental systematics but uses simple background extraction to correct the light curve. The simple hybrid light curves show remarkable agreement with simultaneous ground-based observations, which, however, are not as densely sampled as TESS light curves \citep{2023ApJ...958..188S, 2024ApJ...966..158P}. This method preserves long-term variability. 
The fully hybrid method, on the other hand, considers all systematics rigorously and removes them from the source flux itself. Specifically, the fully hybrid method removes the effect of electronic crosstalk noise and other systematics that are significant in the high-frequency regime, which makes it more suitable for studying rapid variability and quasi-periodicities. However, the full hybrid method, in the process of removing systematics, over-fits long-term variability of the order of the length of a TESS observation sector as demonstrated in Figure 6 of \citet{2023ApJ...958..188S}. As the aim of this work is to analyze the rapid variations in TESS observations, we used fully hybrid light curves for further analysis.

\citet{2024A&A...686A.228O} analyzed the same TESS observations of 3C 371 and included some simultaneous ground-based observations. They used the simple aperture photometric electron flux (SAP\_FLUX) from TESS in their analysis. The simple hybrid light curve obtained using {\tt Quaver} agrees rather well with their light curves as well as with other simultaneous ground-based observations, as demonstrated in \citet{2024ApJ...966..158P}. Since the observations in \citet{2024A&A...686A.228O} best treat the long-variability trends, similar to the simple hybrid light curves in {\tt Quaver}, the variability they found is longer than 7 days for most of the sectors.



Fig.~\ref{fig:flux} shows the TESS Cycle 2 observation of 3C 371 corrected using both simple hybrid and full hybrid methods. The uncertainties on the flux measurements are $\sim$1\% of the mean flux, which reflects the high quality of the TESS data. The whole light curve of Cycle 2 spans sectors 14--26 except for sector 19. Based on different mean flux levels, we divide the light curves into epochs A, B, and C. 
The mean and variance of each sector using the full-hybrid reduction are listed in Table 1. The division into epochs is based on statistical properties, such as the mean being similar in several sectors. 
One reason for selecting this particular division is the presence of a large gap between Sec.\ 18 and Sec.\ 20, which can be effectively used to test our methods for mitigating gaps. Significant variability was noted across the entire baseline of the TESS light curve of 3C 371, which helps in estimating variability timescales using different time-series analysis methods. Hence, the lengthy Cycle 2 TESS observation of 3C 371 is ideal for testing the time-series methods employed in this work.   


Fig.~2 shows the distribution of fluxes observed during each sector and their fits with Gaussian and bimodal distributions, denoted by blue and red curves, respectively. We consider these types of distributions because multi-wavelength observations of AGN have often been seen to follow them \citep{2005MNRAS.359..345U, 2010ApJ...721.1014M, 2018ApJ...857..141S, 2023ApJ...958..188S}. The flux distributions of all sectors show essentially Gaussian profiles. In addition, some sectors (15, 17, 20, and 23) also show some double Gaussian features, indicating that more than one process may be responsible for the observed variability. Sectors 16, 21, 23, 24, and 26 have distributions with high flux tails, indicating that these sectors have ``flares" \citep{2024ApJ...977..166T}, which can also be observed in the light curves (Fig.~\ref{fig:lc})

\section{Data Analysis Methods}\label{sec:carma}
In this section, we will discuss the various time-series analysis methods used in this work. These techniques are PSD, SF, C-ARMA modeling, wavelet decomposition, Bartlett's method, and recurrence analysis. 

\subsection{Continuous-time auto-regressive moving average (C-ARMA) models}
C-ARMA processes have been employed to model the stochastic variability found in many astrophysical sources \citep[e.g.][]{2014ApJ...788...33K, 2017MNRAS.470.3027K}. It has also been used specifically to study the variability of AGNs \citep[e.g.][]{2016A&A...585A.129S, 2017ApJ...834..111C}. The variability observed in AGNs across the electromagnetic spectrum is believed to originate in the outflows, jets, and the accretion disk of the system. The resultant variable emission can interact with various phases of the plasma surrounding the central engine, which can make the observed variability more complex and non-linear in nature. Therefore, including non-linear dynamics in the analysis could provide a better model to explain the observed variability. 

C-ARMA processes directly probe the possible correlation structure present in the observation in the form of a stochastic differential equation which defines the temporal evolution of data. These equations provide higher-order statistics that could characterize specific physical processes better than simple Gaussian processes. Besides, the PSD obtained by these C-ARMA processes can be used to model the complexities arising from a wide variety of phenomena as it is a rational function having multiple auto-regressive (AR) and moving average (MA) coefficients. Some blazars seem to have long-term light curves that are fit better by higher-order C-ARMA processes (such as ($p,q$) = (3,2), (4,3) or (4,1)) than by lower-order ones (e.g. \citet{2018ApJ...863..175G}; \citet{2019ApJ...885...12R}).



In this work, we employed the C-ARMA (2,1) process, commonly known as the Damped Harmonic Oscillator (DHO) which is a second-order differential equation. The motivation behind using this model to examine the variability is the ability of second-order differential equations to explain many physical systems. We note that \citet{2015MNRAS.451.4328K} studied {\it Kepler} AGN light curves and claimed that the simpler C-ARMA (1,0) model, commonly known as Damped Random Walk (DRW), does not do a good job of fitting the variability for all AGNs, especially for studying short-term variability \citep{2014ApJ...788...33K, 2017MNRAS.470.3027K, 2024ApJ...977..166T}. Also, \citet{2022ApJ...936..132Y} pointed out that higher-order ($p,q$) CARMA models are quite susceptible to over-fitting; they also showed how some of the parameters from DHO models can be related to AGN physical properties. 

A C-ARMA (2,1) process $C(t)$ is defined as the solution of the stochastic differential equation \citep{2014ApJ...788...33K}
\begin{multline}
\frac{d^2C(t)}{dt^2} + \alpha_{1}\frac{dC(t)}{dt} + \alpha_0C(t) = \beta_{1}\frac{d\epsilon(t)}{dt} + \beta_{0}\epsilon(t),
\end{multline}
where $\alpha_i$ and $\beta_i$ are the autoregressive and moving average coefficients, respectively. The coefficients are defined such that $\alpha_2$ is equal to unity. Here, $\epsilon$(t) is the Wiener process that provides stochasticity in the behavior of $C(t)$ which is essentially a perturbation around the steady flux state for the observed state $C(t)$.

The left part of the equation determines the evolution of individual perturbations, and the right part determines the shape of the correlated Gaussian noise process. If $\beta_1 = 0$, then the PSD of the noise process follows the behavior of white noise, but it is a type of colored noise when $\beta_1$ is non-zero. So, this equation is essentially equivalent to DHOs with colored noise driving the perturbation, which implies that the astrophysical observation is modeled by several DHOs having fluctuations around a mean flux state. We use the {\tt EzTao} package \citep{2022ApJ...936..132Y} to construct light curves with the DHO model. The given observation is fitted with the C-ARMA (2,1) model, and then the synthetic light curve is produced. This light curve is then used to construct second-order timing products such as PSD and SF.

\subsection{Power spectral density (PSD)}
The stochastic nature of the blazar variability, as manifested in their light curves, usually provides a PSD that can be most simply fit by an inverse power-law. Explicitly, the variability amplitude or the PSD power, $P$, is expressed as a function of the Fourier frequency $\nu$ through $P(\nu) \propto \nu^{\alpha}$. The power spectral index, $\alpha$, describes the nature of the noise. If $\alpha \approx$0, the spectrum is dominated by random white noise. Negative values of $\alpha$ correspond to correlated stochastic noise generally referred to as red noise. 

The Lomb-Scargle periodogram \citep[LSP;][]{1976Ap&SS..39..447L, 1982ApJ...263..835S} is a commonly used technique to estimate the PSD for astronomical observations, as it is sensitive to the gaps in the observations and uses a likelihood optimization method to fit sinusoidal components to the data. The generalized LSP (GLSP) improves on this by fitting the data with both sinusoidal and constant components and also includes the error associated with the flux in the calculation. In our work, we use the GLSP implementation described in \citet{2009A&A...496..577Z} and the routine given by {\tt PYASTRONOMY} \citep[e.g.][]{2021MNRAS.501.5997T, 2023ApJ..running.958..188S, 2024MNRAS.527.9132T}.

The GLSP method models the power spectrum in the frequency domain, but could have some spectral distortions produced by aliasing \citep{2005PhRvE..71f6110K} and red noise leak \citep{1982ApJ...261..337D, 1984ApJ...281..482D}. It is also possible to model the power spectrum directly in the time domain using the C-ARMA process. \citet{2017MNRAS.470.3027K} claimed that the PSD of the observations of space-based telescopes such as {\it Kepler} is well modeled with the DHO process. 


In this work, we used the {\tt EzTao} Python package \citep{2022ApJ...936..132Y} to calculate the PSD of the C-ARMA (2,1) process and then compare it with the GLSP calculations. 

The C-ARMA PSD is then fitted with the broken power law as described in \citet{2024ApJ...977..166T} and is written as 

\begin{equation}
\begin{split}
    P(\nu) &= N \left(\frac{\nu}{\nu_b}\right)^{-\alpha_{high}} + C,~~~ \nu > \nu_b;\\
           &= N \left(\frac{\nu}{\nu_b}\right)^{\alpha_{low}} + C,~~~ \nu < \nu_b~.
\end{split}
\end{equation}

\noindent where $\alpha_{low}$ and $\alpha_{high}$ are the power spectral indices at frequencies lower and higher than the bending frequency $\nu_b$, respectively, with $N$  and $C$ the normalization and instrumental white noise, respectively. We use the modeling module of the {\tt Astropy} Python package to fit the GLSP of a light curve in the frequency range of 0.1-1.0 d$^{-1}$. Lower frequencies do not have enough measurements to properly estimate the parameters, and instrumental noise becomes dominant at higher frequencies. In addition, possible periodicities in the variability of the high-cadence optical light curves are also found in this intermediate frequency range \citep{{2021MNRAS.504.5629R}, 2022Natur.609..265J, 2024MNRAS.528.6608T, 2024ApJ...973...10D}.

\subsection{Structure Function (SF)}
Structure functions can be used to describe AGN variability by quantitatively estimating timescales and providing information on any periodicities present in the data \citep{1978IEEEP..66.1048R, 1997A&A...327..539P, 2014MNRAS.439..703G}. The first-order SF of the process $C(t)$, as described in \citet{1985ApJ...296...46S}, is defined as 
\begin{equation}
    SF(\delta t) = \langle [C(t+\delta t) - C(t)]^2\rangle.
\end{equation}
Thus, the SF quantifies the behavior of variance in flux differences as a function of time lag. The SF measures the variability in the real (temporal) space, which has an advantage over the traditional PSD method where the variability is measured in Fourier (frequency) space and can suffer from various spectral distortions. Any pronounced dip in the SF after a consistent rise with time indicates a variability timescale. If the SF only shows a monotonous increase, any timescale should be longer than the entire length of the observations \citep[e.g.][]{2010ApJ...718..279G}. 



We again used the {\tt EzTao} Python package \citep{2022ApJ...936..132Y} to estimate the SF of the C-ARMA (2,1) process and compare it with that calculated directly from observation.

\subsection{Wavelet Analysis}
While PSD provides insight into the presence of a stochastic process in frequency space, it does not provide any information about how this signal or the stochastic process evolves with time or when that process was present during observation. The temporal evolution is important for studying the consistency of the variability patterns present in the observation. Sometimes, a feature that seems to be very strong in the PSD analysis only lasts for a portion of the observation and should not be considered as indicating that the type of variability is present throughout the data. The duration of variability is also important for its appropriate modeling. 

Analysis in terms of wavelets is a commonly used technique to probe both frequency and time domains simultaneously \citep[e.g.][]{1998BAMS...79...61T}. Wavelets have been used extensively to study variability \citep[e.g.][]{2023MNRAS.522..102R} and to search for periodicity \citep{2018NatCo...9.4599Z, 2020MNRAS.492.5524O, 2020ApJ...896..134P, 2022Natur.609..265J, 2024MNRAS.527.9132T, 2024MNRAS.528.6608T} in multi-wavelength observations of blazars. In this method, the light curve is represented in terms of a wavelet function, which is a function of both frequency and time. An example of such a function is the Morlet wavelet $\phi(\eta)$ described in Eq.~1 of \citet{1998BAMS...79...61T} as

\begin{equation}
    \phi(\eta) = \pi^{-1/4}\exp(-i\omega_0\eta)\exp(-\eta^2/2).
\end{equation}

This Morlet wavelet is a plane wave modulated by a Gaussian and expressed as a function of non-dimensional time parameter $\eta$. Here $\omega_0$ is the non-dimensional frequency whose value is decided such that the wavelet function is admissible, which means that the function must be decomposed in both time and frequency space and have zero mean \citep{1992AnRFM..24..395F}.

The weighted wavelet Z-transform (WWZ) is a transform of the wavelet function used to analyze unevenly sampled data, as it is less sensitive to gaps present in various ground- and space-based observations \citep{2005NPGeo..12..345W}. Recently, this technique has been used to detect quasi-periodicities present in some of those observations \citep{2017ApJ...849....9Z, 2021MNRAS.501.5997T, 2023MNRAS.518.5788O, 2024ApJ...977..166T}. In this work, a publicly available WWZ code\footnote{\url{https://www.aavso.org/software-directory}} has been used. This code uses the Morlet wavelet to fit the given observation.
The average of WWZ over the whole duration of the observations yields the mean WWZ power as a function of frequency, which is essentially a periodogram having a $\chi^2$ distribution with 2 degrees of freedom \citep{1996AJ....112.1709F}. 

\begin{table*}\label{tab:1}
 \caption{The statistical properties of individual sectors of Cycle 2 observation of 3C 371. Also listed are the best-fit parameters of the broken power model used to fit the CARMA PSD of each sector. The corresponding timescales calculated using the SF and GLSP methods are also listed. For Sectors 18, 20, and 21, the timescales corresponding to observational SF are listed, as there are no dips in the C-ARMA SF of these sectors. }
\begin{tabular}{|c|c|c|c|c|c|c|c|c|c|}
\hline
Sector  & mean & $\sigma$& $N$ & $\alpha_{low}$ & $\alpha_{high}$ & C-ARMA $\nu_b$ ($d^{-1}$) & SF & GLSP $\nu_b$  \\ \hline
14      & 1450.9 $\pm$ 0.2  &  4.6 $\pm$ 0.2 & 0.25 $\pm$ 0.03 &-2.4 $\pm$ 0.8 &2.7 $\pm$ 0.3 & 0.14 $\pm$ 0.02 & 0.15 $\pm$ 0.02 & 0.17 $\pm$ 0.03 \\
15      & 1449.4 $\pm$ 0.4 &  6.7 $\pm$ 0.4 & 0.28 $\pm$ 0.04 &-2.4 $\pm$ 0.8 &3.4 $\pm$ 0.5 & 0.21 $\pm$ 0.04 & 0.21 $\pm$ 0.02 & 0.18 $\pm$ 0.04\\
16      & 1449.6 $\pm$ 0.3 & 7.4 $\pm$ 0.3 & 0.43 $\pm$ 0.05 &-2.7 $\pm$ 1.2 &3.2 $\pm$ 0.6 &0.21 $\pm$ 0.03 & 0.22 $\pm$ 0.03 & 0.24 $\pm$ 0.02 \\
17      & 1444.5 $\pm$ 0.4 &  6.6 $\pm$ 0.4 & 0.30 $\pm$ 0.06 &-1.9 $\pm$ 0.6 &2.0 $\pm$ 0.4 &  0.15 $\pm$ 0.01 & 0.20 $\pm$ 0.03 & 0.11 $\pm$ 0.03\\
18      & 1433.4 $\pm$ 0.3 &  8.9 $\pm$ 0.3 & 0.25 $\pm$ 0.04 &0.00 $\pm$ 0.1 & 2.4 $\pm$ 0.5 & 0.17 $\pm$ 0.03 & 0.32 $\pm$ 0.02 & 0.12 $\pm$ 0.03\\
20      & 1426.4 $\pm$ 0.6 &  8.2 $\pm$ 0.7 & 0.12 $\pm$ 0.05 &-0.12 $\pm$ 0.1 & 2.6 $\pm$ 0.7 & 0.29 $\pm$ 0.04 & 0.28 $\pm$ 0.03 & 0.16 $\pm$ 0.02\\
21      & 1425.6 $\pm$ 0.5 & 9.1 $\pm$ 0.5 & 0.14 $\pm$ 0.04 &0.23 $\pm$ 0.02 &1.8 $\pm$ 0.3 &0.19 $\pm$ 0.02 & 0.18 $\pm$ 0.02 & 0.16 $\pm$ 0.03\\
22      & 1417.8 $\pm$ 0.5 &  8.5 $\pm$ 0.4 & 0.78 $\pm$ 0.11 &-3.4 $\pm$ 0.6 & 3.9 $\pm$ 1.2 & 0.32 $\pm$ 0.04 & 0.24 $\pm$ 0.03 & 0.24 $\pm$ 0.04\\
23      & 1430.5 $\pm$ 0.7 &  11.5 $\pm$ 0.7 & 0.37 $\pm$ 0.05 &-2.8 $\pm$ 0.4 &3.4 $\pm$ 0.4 & 0.13 $\pm$ 0.01 & 0.14 $\pm$ 0.02 & 0.10 $\pm$ 0.03\\
24      & 1460.1 $\pm$ 0.5 &   8.9 $\pm$ 0.4 & 0.18 $\pm$ 0.02 &-0.7 $\pm$ 0.2 &3.4 $\pm$ 0.6 & 0.28 $\pm$ 0.01 & 0.28 $\pm$ 0.02 & 0.13 $\pm$ 0.04\\
25      & 1452.4 $\pm$ 0.4 &  9.2 $\pm$ 0.3 & 0.17 $\pm$ 0.03 &-0.7 $\pm$ 0.3 &3.9 $\pm$ 0.8 &0.30 $\pm$ 0.03 & 0.29 $\pm$ 0.03 & 0.28 $\pm$ 0.04\\
26      & 1457.0 $\pm$ 0.2 &  6.6 $\pm$ 0.2 & 0.15 $\pm$ 0.01 &-1.2 $\pm$ 0.3 &1.8 $\pm$ 0.4 & 0.14 $\pm$ 0.02 & 0.14 $\pm$ 0.01 & 0.09 $\pm$ 0.03\\  \hline
\end{tabular}

\end{table*}

\subsection{Bartlett's Method}

Bartlett's method \citep{1948Natur.161..686B} is an approach to computing the periodogram of a time series by averaging the periodogram of equal segments of the light curve, resulting in smoothing of the periodogram by reducing the variance. This would result in the reduction of noisy features. In this method, the power is averaged in the Fourier space at the same frequency for a particular light curve. First, the light curve is divided into $K$ segments with the same duration (or the same frequency resolution in the Fourier space). Then a Fourier transform (FT) is calculated for each segment,  and the subsequent periodogram is then averaged over all $K$ segments \citep{2022arXiv220907954B}. 

One of the most common issues encountered with the analysis of multi-sector TESS observations is how to perform careful treatment of the gaps between the sectors and also, sometimes, within the sector. The multiple-sector observation can be treated as a time series comprising many sectors separated by gaps. The average periodogram $\overline{P(f)}$ for the $K$ TESS sectors of an observation can simply be written  as 
\begin{equation}
   \overline{P(f)}  = \frac{1}{K}\sum_{i=0}^{K-1} P_i(f)~,
\end{equation}
where $P_i(f)$ is the PSD of the $i^{th}$ sector of the total of $K$ sectors of observations during a TESS Cycle.

\subsection{De-noising the signal}
The observations of AGNs, irrespective of the waveband, primarily consist of stochastic variability usually characterized as red noise, which typically follows an inverse power-law dependence with frequency. The observations also contain some random, uncorrelated, white noise, which is completely independent of frequency. In the Fourier log-log space,  red noise fluctuations appear as a straight line with a negative slope, and the white noise as a constant. This white noise can play a very important role as it defines the base level (or normalization) required to model the red noise for a power spectrum. Therefore, it is important to de-noise the signal for estimating the appropriate significance relative to the red noise or the stochastic variability. 

Wavelet transforms can assist in removing the noise present in the data which could potentially affect our estimation of statistical properties related to an observation. First, we computed the discrete wavelet transform (DWT) for the observation using the {\tt pywt} python package. With this approach, the desired noise level can be specified, and one can then keep the DWT coefficients higher than that value. After applying the desired threshold, an inverse DWT can be applied to retrieve the signal. This technique removes the white noise present at higher frequencies and also increases the power of the PSD at lower frequencies \citep[see][and references therein]{2001C}. 

We can also reduce the white noise present in the data by combining periodograms, i.e., using Bartlett's method. Combining PSDs from various sectors averages the random, uncorrelated, white noise present in the observation across all sectors. We discuss this de-noising technique further in Section \ref{sec:res}.  

\subsection{Recurrence analysis}

Recurrence analysis is a time series analysis technique that can be used to quantify the non-linearity present in the observation and identify the source of variability as deterministic or stochastic \citep{2007PhR...438..237M}. While the auto-correlation function (ACF) estimates the correlations in the amplitudes of a scalar time series, recurrence analysis computes the recurrences of a dynamical system to the same region of its higher-dimensional phase space. An example of the phase space of a dynamical system is angular position versus angular velocity for a pendulum: every point in this phase space represents a solution to the equations of motion for the pendulum, and a series of points, called a ``trajectory", represents how those solutions evolve over time. For a simple pendulum, the trajectory would be a circle. The phase space for a single observable (e.g., if you only have the angular position for the pendulum) can be estimated through a variety of numerical methods. For example, angular velocity can be estimated through a delayed copy of angular position for a pendulum, since the derivative (angular velocity) is merely a phase shift of angular position, like the relationship between sine and cosine. This method is called the Time Delay Method \citep{1981LNM...898..366T} and has been shown to be an equivalent representation of the true phase space \citep{2009Chaos..19b3104R}. We will use the Time Delay method to perform recurrence analysis for all the light curves in this paper. In this method, the choice of time delay used to construct the additional dimensions of the phase space embedding is chosen such that each point is minimally correlated, i.e., is set as the value when the autocorrelation drops to 0. In this way, we are constructing a phase space using data points that are not redundant to each other \citep{2020MNRAS.497.3418P}. 

A recurrence plot (RP) is constructed by comparing every position, or ``state", in phase space to every other one \citep{1987EL......4..973E}. Since each point in phase space corresponds to a point in time, the two axes in the resulting two-dimensional RP correspond to time and the matrix contains entries of 1 (black) and 0 (white) corresponding to recurring and non-recurring states, respectively. Mathematically, if $\vec{x}(t)$ is the phase space trajectory of a system, then the RP of a state occurring at time $i$ recurring at a different time $j$ is expressed as 

\begin{equation}
    R_{i,j} = \Theta (\epsilon_i - ||\vec{x_i} - \vec{x_j} ||), \vec{x_i} \in \mathbb{R}^m,
    i,j = 1,2,..N,
\end{equation}
where $\Theta$ is the Heaviside function, $N$ is the number of considered states $x_i$, $\epsilon$ is a threshold distance used to define proximity in phase space, and $||.||$ is the Euclidean norm \citep{2007PhR...438..237M}. If the distance between the two points in a time series after embedding in the phase space is less than the specified threshold, $\epsilon$, the recurrence matrix element is taken as 1; if the distance is larger, the matrix entry is zero.

The recurrences in phase space are fundamental features of a dynamical system: random processes do not have deterministic equations of motion, so the positions and subsequent correlations in phase space are random. On the other hand, the presence of recurrences (most often seen as diagonal-line features in an RP), signal the presence of possible structured variability, such as (quasi-) periodicity, non-linearity, or even chaos. High-order stochastic systems, like first- and second-order auto-regressive processes, have some amount of memory, which can manifest as patterns in the RP. The usefulness of the RP is based on the critical property that each class of dynamical system has a unique pattern in phase space (e.g., a circle for a simple pendulum versus randomly distributed points for white noise), which allows one to identify different classes of variability in real data. \citep[For details, see][and references therein]{1987EL......4..973E, 2020MNRAS.497.3418P}.

 
If one were to sum along the diagonals of the 2-D recurrence matrix, you would retrieve an analog to an autocorrelation function (ACF). In an ACF, the first peak gives an indication of a variability timescale, and any subsequent peaks can indicate a possible periodicity. Similarly, the structures in a RP around the diagonal will also give indications about the presence of variability. The patterns observed in an  RP can be used to detect structures present in the data. For instance, a uniformly distributed RP indicates that the data is purely random. Any periodic or quasi-periodic features are characterized by repeating structures (diagonal lines, checkerboard features). For non-stationary data, one might observe changes in the patterns across the RP \citep{2007PhR...438..237M}, while vertical or horizontal lines correspond to a stationary or very slowly changing dynamical state. 

Recently, \citet{2020MNRAS.497.3418P} performed an RA of a {\it Kepler} observation of the AGN KIC 9650712, which was put forward as a QPO candidate in \citet{2018ApJ...857..141S}. The timescale of the QPO was confirmed using RA, and its RP was distinct from the RP of stochastic surrogates (simulated light curves with the same power spectrum and flux distribution), supporting the claim that the source of the QPO is a deterministic, possibly nonlinear, process.


\section{Results}\label{sec:res}

In this section, we will discuss the results for each sector individually using PSD, SF, and RA. We also present the results for each epoch as well as for the entirety of the Cycle 2 TESS observations of 3C 371. We compare the results using different methods to mitigate gaps (C-ARMA and Bartlett's method) and include de-noising of the observations via wavelet decomposition. We also discuss the significant variability observed in the epoch-wise light curves using WWZ and GLSP methods. 
\subsection{Sector-wise analysis}

\begin{figure*}
\hspace{-1.0cm}\includegraphics[scale=0.4]{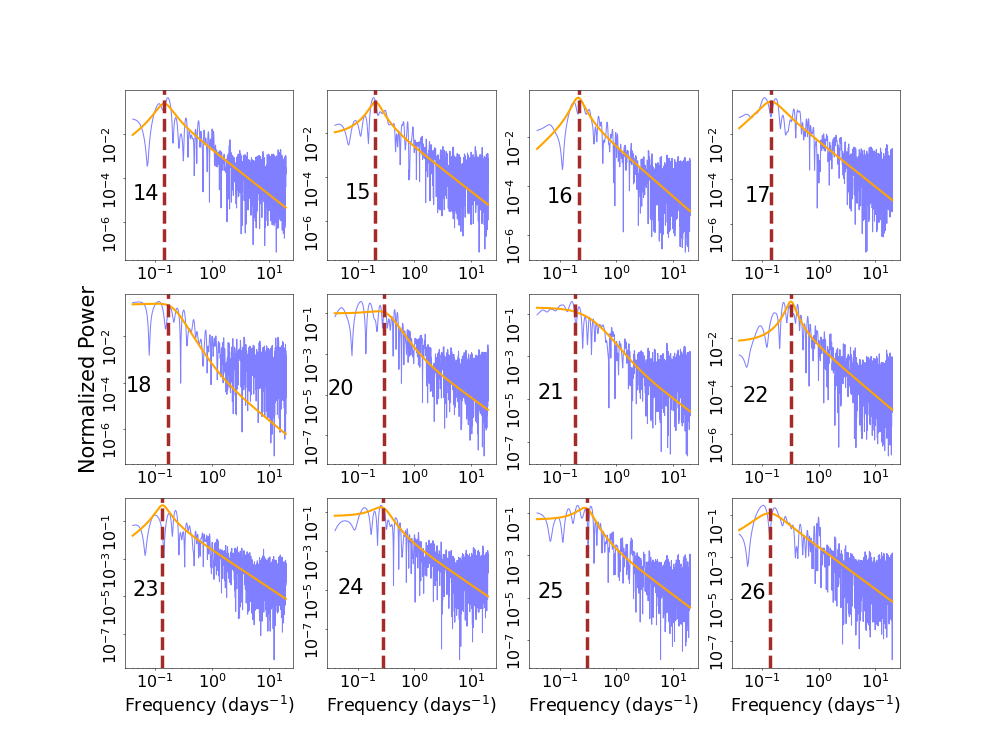}
\caption{Periodogram estimations for each sector of the Cycle 2 TESS observations of 3C 371. The blue and orange curves correspond to the Lomb-Scargle periodogram and the periodogram obtained from C-ARMA modeling respectively. The vertical dashed lines mark the bending frequencies $\nu_b$ where the  spectral index changes, corresponding to a variability timescale present in the observation.}
\end{figure*}\label{fig:psd}

Fig.~3 presents the PSD analysis for the individual sectors. The blue curve shows the PSD calculated by the GLSP method. The orange curve is the PSD calculated by the C-ARMA (2,1) DHO process. The best-fitting parameters obtained by fitting the sectors individually with such broken power laws are listed in Table 1. The vertical dashed curves in Fig.~3 denote the bending frequency where the behavior of the spectral index $\alpha$ changes and are also the indicator of the variability timescales associated with the emission region and the cooling timescales \citep{2002MNRAS.332..231U, 2015MNRAS.451.4328K, 2021ApJ...909...39G,2024ApJ...973...10D}. The timescales vary from 0.13 to 0.32 days$^{-1}$ (d$^{-1}$ hereafter) corresponding to the order of 3--7 days. The values of $\alpha_{high}$ for various sectors lie between 1.8 and 3.9, and except for sectors 22 and 25, they have $\alpha_{high}$ between 2 and 3 (within errors), which is commonly observed for blazars. Sectors 22 and 25 also have the shortest variability timescales (or the highest $\nu_b$ values).  These higher values of the spectral index imply that there could exist processes different from red noise that affect the fluxes and variability timescales. Table 1 also lists the timescales obtained with the GLSP analysis. These timescales are largely in agreement with those obtained for C-ARMA, within errors. The plausible reason for their differences is the presence of multiple peaks with similar power near the bending frequency, and C-ARMA equations smooth these features when obtaining the solution for these observations. The SF timescales also agree with the GLSP timescales for most sectors. For Sector 24, the GLSP timescale differs significantly from the C-ARMA and SF ones. In the GLSP of this sector, there are peaks of similar power at frequencies of 0.13 d$^{-1}$ and 0.28 d$^{-1}$, and it is possible that C-ARMA analysis picked one of them, which would better fit the observation.  In the SF plots, the first dip (at 3.5 days) is less pronounced than the one at 7 days. So, there are features present at both frequencies. There could be a similar situation in sectors 18 and 20, where no dips are detected in C-ARMA SF.  

\begin{figure*}
\hspace{-1.cm}\includegraphics[scale=0.4]{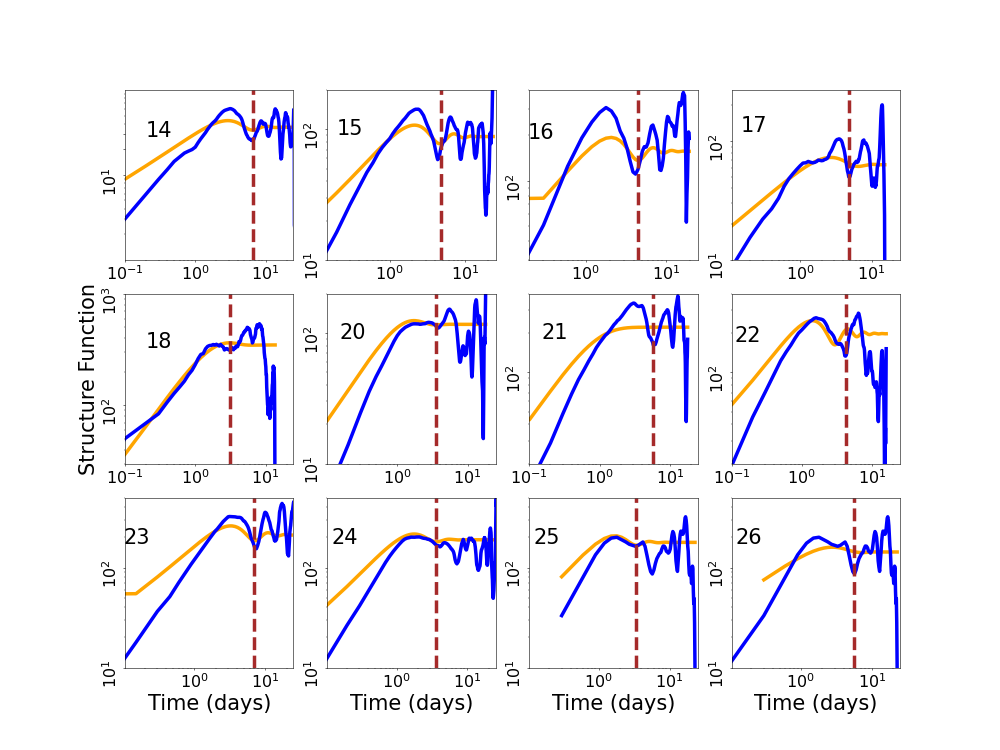}\\
\caption{Structure Function (SF) calculations for each sector (numbered in each panel) of Cycle 2 TESS observations of 3C 371. The blue curve is the SF curve estimated directly from the observation and the orange curve corresponds to the SF calculated by C-ARMA modeling. The first dip in the SF corresponds to the timescale of variability present in the observation and is denoted by the {vertical dashed line.} }
\end{figure*}\label{fig:sf}

Fig.~4 shows the SF result for each sector of the Cycle 2 observations. The blue and orange curves represent the SF calculated using observation and the C-ARMA DHO process, respectively. The vertical dashed lines show the first dips in the SF (observations) which, as in Fig.~3,  indicate the variability timescales. These variability timescales for the individual sectors lie between 3.0 and 7.0 days,  in agreement with the range found from the C-ARMA PSD analysis. For most of the sectors, the first dip in the observational SF is consistent with that in the C-ARMA SF. For Sectors 18, 20, and 21, there are no dips in C-ARMA SFs, whereas they are seen via the observational SFs. If a higher-order C-ARMA process were to be considered, it is possible that additional features in the C-ARMA SF would also be visible and which could  be consistent with the dips found in the observational SF. For Sector 17, the discrepancy in observational and C-ARMA SF could be due to the uneven shape of the observational SF. So, the exact dip of the function is very difficult to estimate. For sector 18, it is interesting to note that the bending timescale corresponds to the period of the second dip in the structure function. However, for consistency, the result corresponding to only the first dip is included in this work. 

\begin{figure*}
\includegraphics[scale=0.42]{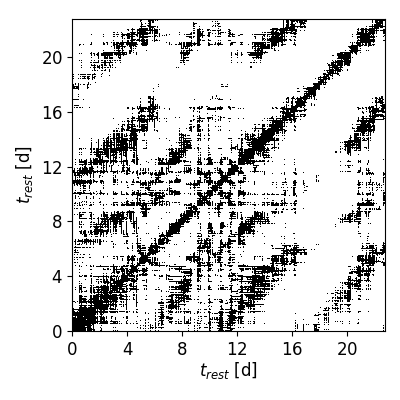}\includegraphics[scale=0.42]{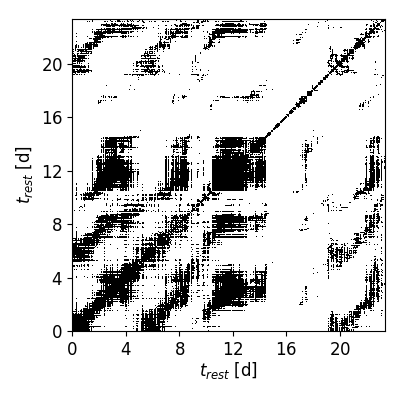}\includegraphics[scale=0.42]{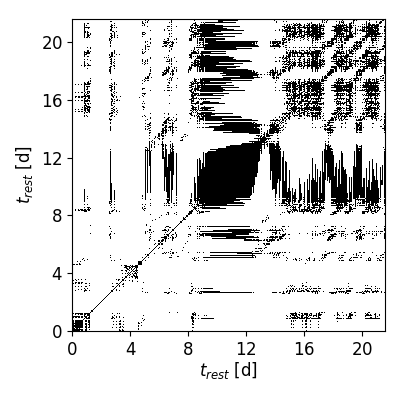}\includegraphics[scale=0.42]{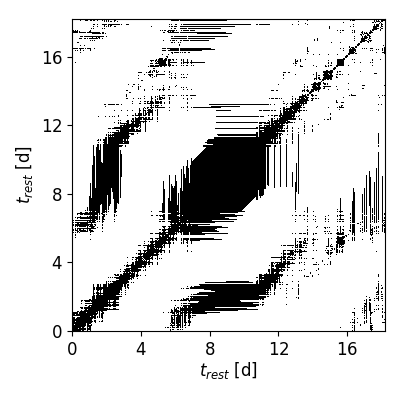}\\
\includegraphics[scale=0.42]{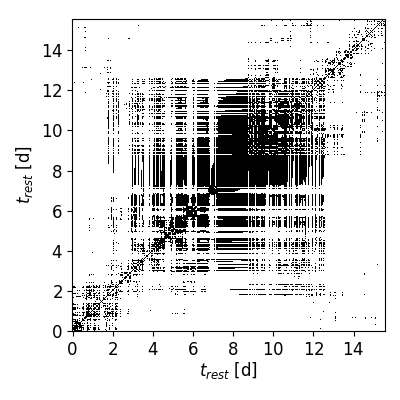}\includegraphics[scale=0.42]{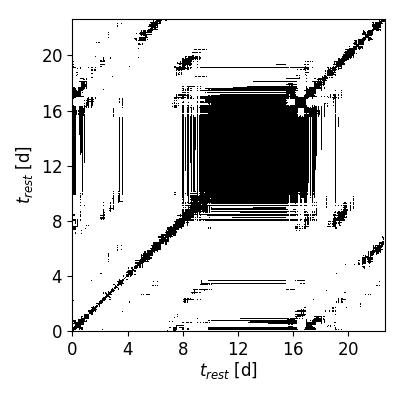}\includegraphics[scale=0.42]{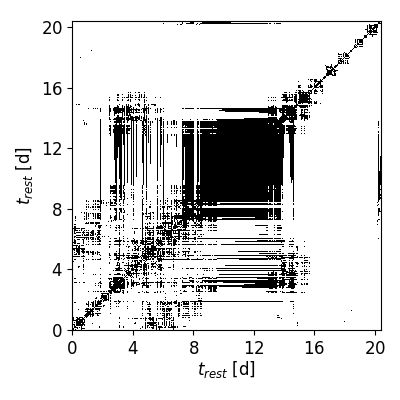}\includegraphics[scale=0.42]{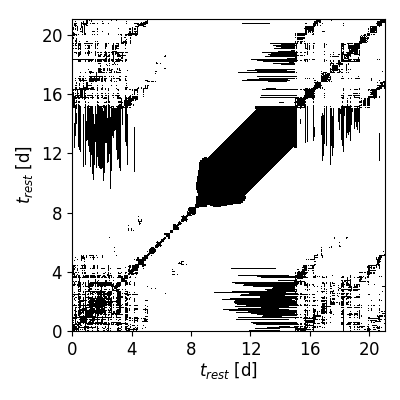}
\includegraphics[scale=0.42]{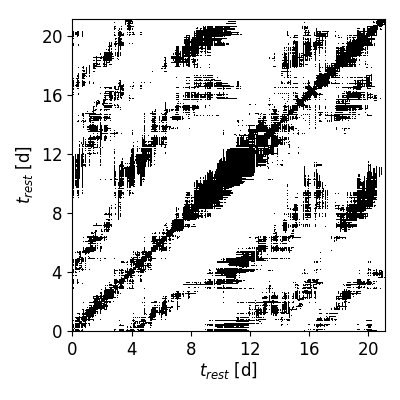}\includegraphics[scale=0.42]{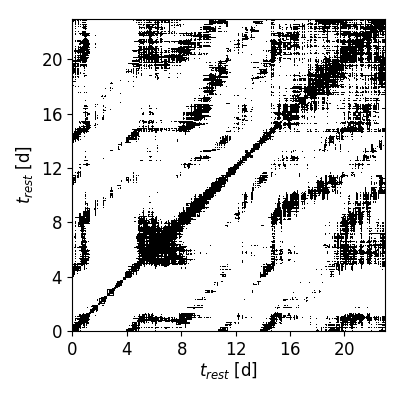}\includegraphics[scale=0.42]{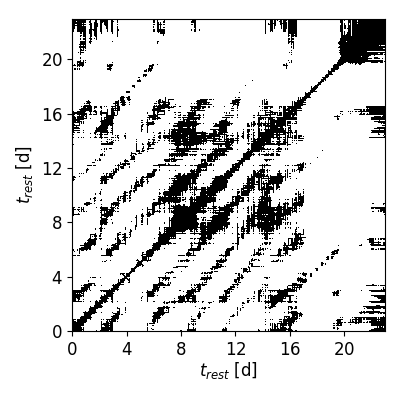}\includegraphics[scale=0.42]{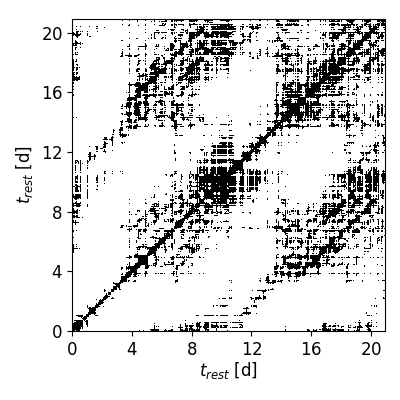}
\caption{Recurrence plots of the individual sectors of Cycle 2 observations, in the order 14, 15, 16, 17 (top row), 18, 20, 21, 22 (middle row), 23, 24, 25, and 26 (bottom row). The black dots in the plots correspond to the value of 1 in the recurrence matrix whereas the white dots correspond to a zero entry in the recurrence matrix. 
}   
\end{figure*}\label{fig:sf}

Fig.~5 presents the RP for all sectors of the Cycle 2 observation of 3C 371. The black dots correspond to an entry of 1, where the distance between the two points in phase space is less than the given threshold. Otherwise, the entry in the recurrence matrix is zero. Note that the RP is symmetric about the main diagonal. The quantity $t_{rest}$ corresponds to the days since the start of the light curve observation. For example, in the first RP shown in Fig.~5, we see the position on the figure at $x=12d$ and $y=2d$ contains an entry of 1 (black dot), indicating that the two points in the light curve at 2 days and 12 days from its start are within close proximity of each other in the phase space embedding of the light curve and thus represent a recurrence. The points at $x=13d$ and $y=3d$ are also recurrent and continue to be so for several days, resulting in a diagonal line in the RP of length 4 days. The entries surrounding these positions are also recurrent, resulting in a diagonal structure that is somewhat broadened. We see a similar pattern throughout the RP, indicating recurrences throughout the light curve. In fact, none of these RPs throughout Fig.~5 are merely uniform, signifying that there is more than pure randomness in the data. Repeating structures such as diagonal lines and checkerboard features are seen in every RP, signaling the presence of processes that drive the observed variability. The repeated structures along the diagonal line in sectors 14, 15, 17, and 23--26 indicate the presence of structured variability or quasi-periodicity. In sector 16, which follows the tailed distribution, there are distinct checkerboard features that imply that there is another process, such as ``flares" that affects the variability present in the observation. The RPs for sectors 18, 20, and 21 have distinct vertical and horizontal lines, implying that the state of the system changes very slowly. 

Regarding the inconsistent dips between the C-ARMA SFs and observational SFs in sectors 18--21, C-ARMA assumes the time series is a stationary process and so does not recover nonlinearity, while non-stationarity is at least somewhat accounted for by our investigating the light curve sector-by-sector. So, merely fitting a higher-order C-ARMA process may not be sufficient (and may actually be inappropriate) as it still can only recover variability due to linear correlations. Another issue could be the inability of the DHO model to model the light curve efficiently and consequently be inadequate to provide a proper model of the variability. This inconsistency in dip locations produced by either of those reasons could be manifested in the RA of these observations. The vertical and horizontal lines in the RPs imply that changes in the dynamics of the system are slow. There may be a nonlinear process going on, or non-stationarity in higher-order statistics (beyond just the mean) that is not accounted for in the sector-by-sector approach. Interestingly, these sectors also show a common block-like feature in the RPs, which implies that the nature of variability shown by these sectors is different from that of other sectors. Thus, methods using different approaches (SF, RA) could be helpful to understand and handle possible inconsistencies found in the analyses 
and can be particularly effective in characterizing the variability in these light curves.

\begin{figure*}
\hspace{-1cm}\includegraphics[scale=0.4]{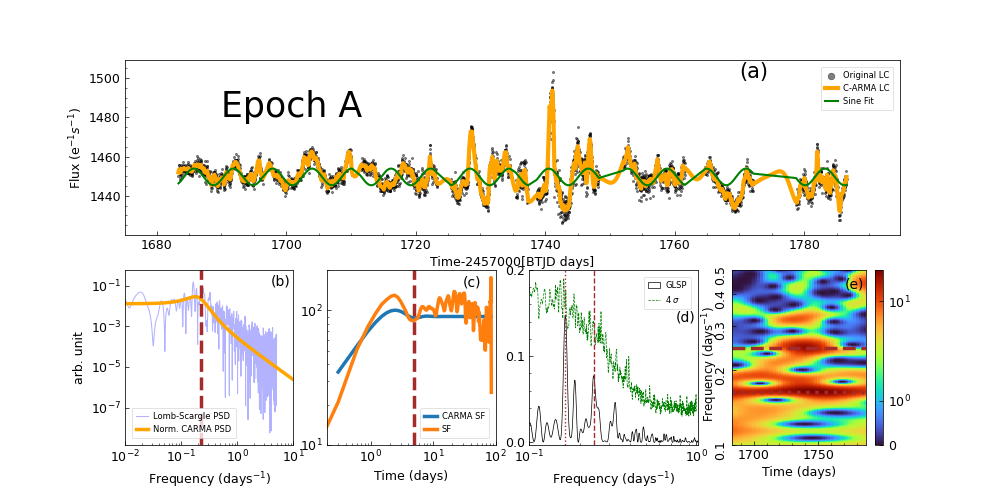}
\caption{Timing analysis results for epoch A. The light curve for epoch A is plotted in panel (a) along with the C-ARMA generated light curve and a sine fit with a period of 6 days. Panels (b) and (c) respectively show the PSD and SF calculated directly from the observations and from the C-ARMA fit. Panel (d) shows the GLSP results along with 4 $\sigma$ significance curve. The brown dotted and brown dashed curves respectively, correspond to a possible quasi-periodic and a variability frequency. The corresponding WWZ color-color diagram is plotted in panel (e). The QPO and variability frequencies are denoted by black dotted and brown dashed curves, respectively.}
\end{figure*}\label{fig:epocha}

\begin{figure*}
\hspace{-1cm}\includegraphics[scale=0.4]{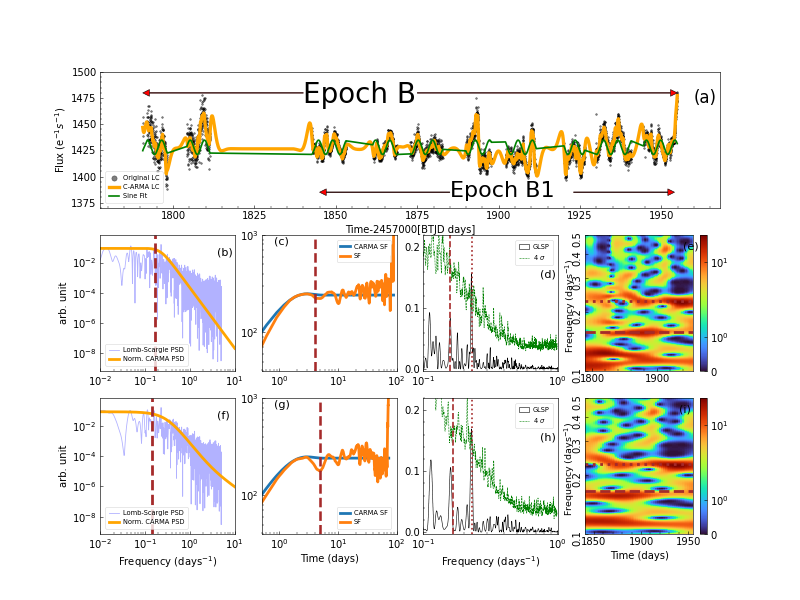}
\caption{Same as in Figure 6 but for epochs B (center panels) and B1 (lower panels).}
\end{figure*}\label{fig:epochb}

\begin{figure*}
\hspace{-1cm}\includegraphics[scale=0.4]{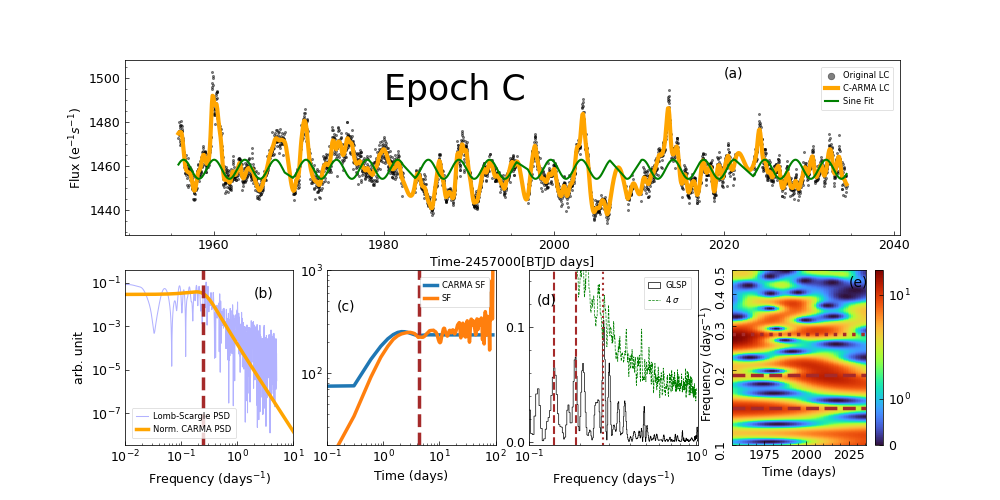}
\caption{Same as in Figure 6 but for epoch C. }
\end{figure*}\label{fig:epochc}

\begin{figure*}
\hspace{-1cm}\includegraphics[scale=0.4]{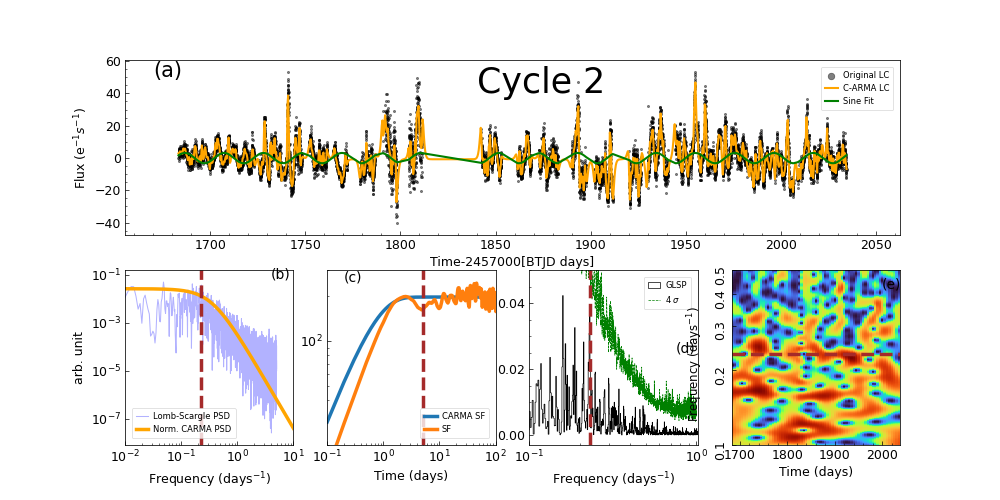}
\caption{Same as in Figure 6 but for whole Cycle 2 observation. }
\end{figure*}\label{fig:epochabc}

\subsection{Epoch-wise analysis}

Fig.~6 shows the light curve and analysis results for epoch A. The upper panel shows the original light curve and the one generated using the C-ARMA DHO process. The C-ARMA light curve predicts the data between the gaps and computes the corresponding PSD and SF which are plotted in panels (b) and (c). The bending frequency $\nu_b$ for epoch A is found to be 0.21 d$^{-1}$, corresponding to $\approx$ 4.8 days; $\alpha_{high}$ is  $2.8\pm 0.5$,  consistent with those found for individual sectors. The first dip in the SF plot is visible for both observational and C-ARMA SF and is found at $\sim 4.6$ days, consistent with that found by PSD analysis. 

Additionally, the GLSP of epoch A shows a strong signal at around 6 days which could be interpreted as possible quasi-periodic oscillations (QPOs). 
This is supported by the SF analysis, which shows multiple dips and peaks. 
To assess the presence of quasi-periodicity, we estimated the significance of the peak in the GLSP as described in \citet{2024ApJ...977..166T} which implements the method explained in \citet{2013MNRAS.433..907E}. In this method, 10,000 light curves are simulated with the PSD and PDF obtained from the observation and then the level of significance at a particular frequency is derived from the fraction of simulations that exceed the observed value. \citep[For more details, see][]{2024MNRAS.528.6608T, 2024MNRAS.527.9132T}. A given signal will only be considered quasi-periodic if its significance is at least 99.99\% (4$\sigma$). 

Panel (d) shows the generalized Lomb-Scargle periodogram and the 4$\sigma$ confidence curve, plotted in black and green curves, respectively. The peak at $\approx$ 6 days, marked by a brown dotted curve, is found to be at least 4$\sigma$ significant. The dashed brown curve shows the variability timescale at around 4.8 days estimated by the PSD and SF methods. There is also a signal found at this frequency but its significance is less than that found at around 6 days. These two signals can also be found in the wavelet color-color diagram, plotted in panel (e). The variability timescale, denoted by the dashed black line, is weak at the beginning of the epoch. This signal strengthens around the middle of Sec.\ 15 and becomes weak again in the middle of Sec.\ 17. The claimed QPO signal, on the other hand, is found to be persistent throughout all of Epoch A. This suggests that the variability timescale shown in the PSD and SF analysis is introduced by a process different from the one that produces the claimed QPO. This other process could be related to the flare-like structures in the middle of the light curve which also makes it noisier. The upper panel also shows the light curve fit with the sinusoidal function with the nominal QPO period, plotted in green. The temporal baseline of the observation is 102 days, corresponding to 17 Cycles of a 6 day QPO. 

The bottom panel of Fig.~7 shows the time-series results for epoch B which comprises sectors 18--23, excluding sector 19. This epoch is of special interest as it has a huge gap due to the absence of data in Sec.\ 19, which alone constitutes more than 1/6th of the whole epoch. For epochs A and C, the total of any gaps never exceed 10\% of the full baselines, but for epoch B, the gaps exceed 20\% of the baseline, which could adversely affect the significance estimation, as shown through simulations in \citet{2024MNRAS.528.6608T}. The light curve generated by the C-ARMA DHO process seems to plausibly bridge the gaps between the observations, as is seen for the other two epochs. However, the process only yields a straight line for the large gap with no data in Sec.\ 19, indicating that the bounds of the prediction for a long gap are limited by the mean for sparse data.


Initially, we estimate the PSD for the C-ARMA light curve and plot it in panel (b); $\alpha_{low}$ and $\alpha_{high}$ are found to be $0.10 \pm 0.12$ and $2.3 \pm 0.2$, respectively. The bending frequency is $0.17\pm 0.02$ d$^{-1}$,  similar to that found for both epoch A and epoch C. In panel (c), there is a difference between  the locations of the first dip for the observation and C-ARMA SFs. The first dip for the observational SF is 4.0 d, whereas for C-ARMA SF, it is 5.0 d; this is unlike for epochs A and C, where the locations of dips found with observational and C-ARMA SF are consistent. Several dips and peaks can be seen in the SF of this epoch, which can indicate the possible presence of quasi-periodicity. We note that the GLSP and WWZ  methods are commonly used to analyze uneven and irregular light curves, so there is reason to expect that they will not be greatly affected by the large gap. Panel (d) shows the GLSP curve and the 4$\sigma$ confidence level (in green). The dotted brown curve corresponds to the signal at 0.23 d$^{-1}$ and has a significance of more than 4$\sigma$, signaling the presence of a QPO at that frequency. The dashed brown curve shows the bending frequency at 0.17 $d^{-1}$, providing the variability timescale. In the wavelet color-color diagram in panel (e), the claimed QPO signal appears to be persistent throughout the observation aside from during the gap, further supporting the reality of   this QPO signal. The wavelet diagram does, however, exhibit considerable noise, and some weak signals may arise from the huge gap in the light curve. A sinusoid, with a period of 4.38 d, is plotted along with the light curve of Epoch B in the upper panel in green; it is fitted only when data are present. The total duration of epoch B is 164.1 days, encompassing 38 Cycles of that apparent QPO period.

The QPO signal found in epoch B appears to be highly significant. However, the lengthy gap within epoch B might inflate the significance of the peak. Also, the variability timescales calculated through different methods are not consistent. So, to better estimate the significance of the QPO signal and to see if there is a consistent variability timescale, we eliminate Sectors 18 and 19 from epoch B, resulting in epoch B1, shown in panel (a) of Figure 7. Epoch B1 comprises sectors 20--23, which altogether have gaps amounting to less than 5\% of the total temporal baseline and so now can be treated like epochs A and C. In the bottom row of Fig.~7, we show the results for epoch B1. In panel (f), we plotted the C-ARMA PSD along with the GLSP and also marked the bending frequency at 0.14 $d^{-1}$,  which is slightly lower than what is found for epoch B; 
$\alpha_{low}$ and $\alpha_{high}$ are found to be $0.08 \pm 0.1$ and $1.6 \pm 0.2$, which also differ. In the SF calculation,  the values obtained directly from the light curve (5.0 d) and the C-ARMA process (5.95 d) are not consistent, as for the entire epoch B. However, the first dip in the observational SF for epoch B1 is sharper than that of epoch B, where it is wide, possibly due to the huge gap. A strong peak in the GLSP (panel (h)), with more than $4\sigma$ confidence, is found at the frequency of 0.23 d$^{-1}$, similar to epoch B. There is also a peak at the variability timescale estimated by the C-ARMA process. The WWZ color diagram, plotted in panel (i), is less noisy when compared to that of epoch B and the signal at 0.23 d$^{-1}$ is persistent throughout the observation.  The variability timescale estimated by the C-ARMA process was also found to be strong in the first half of epoch B1, but it slowly weakens while the QPO frequency gains strength during the same duration. The variability timescale predicted by the C-ARMA process is also found in both the GLSP and WWZ analyses.  We see that the presence of a large gap could modestly affect the variability timescales but is less likely to affect the quasi-periodicity, owing to the ability of these methods to carefully treat irregular data, like those from TESS. The period of 4.3 days corresponds to 26 Cycles in the temporal baseline of 112.5 days for epoch B1.

Fig.~8 shows the light curve and time-series analyses results for epoch C. The upper panel shows the original light curve and the one generated by the C-ARMA process. The C-ARMA light curve models the light curve well and also models the gaps within. The PSD of the light curve generated by the C-ARMA DHO process yields $\alpha_{low}$, $\alpha_{high}$, and bending frequency to be  $0.0\pm 0.3$, $2.8\pm 0.5$, and $0.26\pm 0.03$d$^{-1}$, respectively. The near-zero value of $\alpha_{low}$ suggests that the fit follows the bending power law model. The value of $\alpha_{high}$ is found to be consistent with its measurements made individually in each sector. The SF for epoch C is plotted in panel (c), where the first dip for both observational and C-ARMA SF is seen to be consistent and has the value of $4.26\pm 0.3$ days, which is also consistent within errors with that estimated using PSD analysis. Like epochs A and B, the SF of epoch C displays several dips and peaks which could signal the presence of quasi-periodicity, the presence of which is examined in the next two panels of Fig.~8. 

Panel (d) plots the GLSP along with the 4$\sigma$ significance curve. A signal at around 3.6 days is found to be at least 4$\sigma$ significant and is also found to be persistent throughout the observation except after the middle of sector 26 in the wavelet color-color diagram plotted in panel (e). The variability timescale and the quasi-period are consistent with the measurement errors; hence, we do not mark the variability timescale separately in the wavelet plot. The concentration of WWZ power throughout the whole epoch is also found at other frequencies which are marked by dashed black lines in panel (e). The signal at 0.19 d$^{-1}$ is weak in the first half of the epoch but gradually becomes strong in the second half and also appears as a peak in the GLSP, although with less significance than the QPO frequency. However, the evolution of the signal at 0.14 d$^{-1}$ is the opposite, i.e., it is strong at the beginning of the epoch and then weakens. A possible implication is that these two variability timescales were introduced to the light curves through different physical processes and the strength of these processes changes during the epoch. However, the similar temporal evolution of the signal at 0.14 d$^{-1}$ and QPO frequency 0.28 d$^{-1}$ suggests that they are introduced by physical processes of similar nature and we note  these two frequencies can well be harmonics. 
The duration of the epoch C is 78.6 days which corresponds to $\approx$ 22 Cycles of the claimed QPO period of 3.6 d. The green curve in the upper panel shows a sinusoidal fit to the light curve with the period of the claimed QPO signal. 

\subsection{Cycle 2 Analysis}
We also performed an analysis of the entire Cycle 2 observations, including all 12 sectors;  the results are plotted in Fig.~9. For this analysis the Cycle 2 observations have been converted into a time series with zero mean. We then fit the resultant light curve with the C-ARMA DHO process and the result is over-plotted in the upper panel of Fig.~9. The C-ARMA PSD yields values of $\alpha_{low}$ and $\alpha_{high}$ to be, respectively, $0.08\pm 0.2$ and $2.2\pm 0.4$. The bending frequency obtained for the whole Cycle 2 observation is 0.16 d$^{-1}$ which corresponds to a variability timescale of 6.25 d. The C-ARMA SF does not exhibit any dip while the first dip in the observational SF is found to be around 5 days. The GLSP plot in panel (d) has many more peaks in the red-noise regime (0.1--1.0 d$^{-1}$) when compared to the periodograms of individual epochs. Here, no peak has 4$\sigma$ confidence nor are there any strong indications of extended signals in the wavelet diagram in panel (e). 

\begin{figure*}
\includegraphics[scale=0.8]{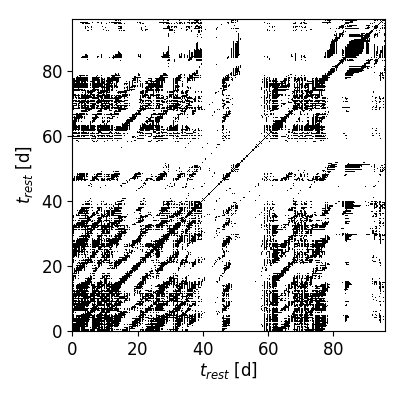}\includegraphics[scale=0.8]{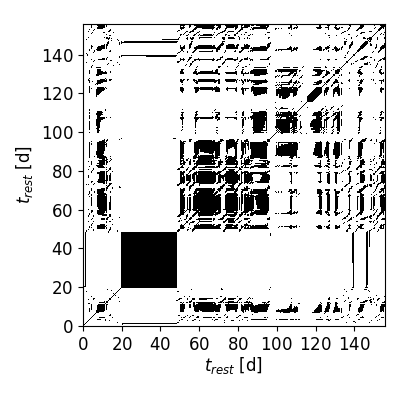} \includegraphics[scale=0.8]{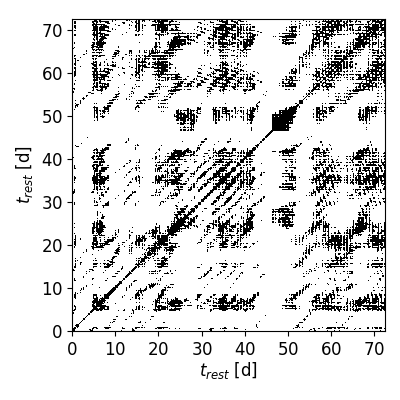}\includegraphics[scale=0.8]{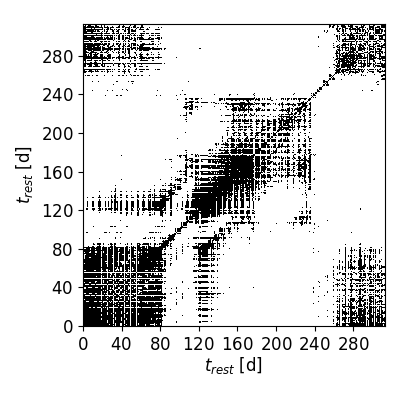}
\caption{ Recurrence plots (RPs) for all 3 epochs and the entirety of Cycle 2 (bottom right panel) observations. The top left, top right, and bottom left panels correspond to RPs of epochs A, B, and C respectively. The diagonal lines and checkerboard structures in the plots signal the presence of possible quasi-periodicity. }
\end{figure*}\label{fig:rae}

\subsection{Recurrence Analysis}
Fig.~10 shows the recurrence plots for all three individual epochs as well as the entire Cycle 2 observation. 
For epochs A and C, checkerboard structures and diagonal lines can be seen clearly, which are patterns frequently observed when quasi-periodicity is present in the data \citep{2020MNRAS.497.3418P}. Despite having a huge gap, the RP of epoch B shows repeated checkerboard features, implying that the recurrence analysis can be usefully employed to display variability for the data even with big gaps. In the full Cycle 2 RP there are some recurring features in the RP, such as aperiodically distributed vertical and horizontal lines, that are not as prominent as those seen in the RPs of individual epochs, possibly due to its non-stationary nature. Besides, these features are random and distinct from the checkerboard structures signaling quasi-periodic variability. The light curve of the Cycle 2 observation, plotted in Fig.~9, consists of many flare-like structures that could affect the variability properties manifested in this light curve. To carefully assess this effect, we will use two methods in the next sub-sections.

\begin{figure*}
\includegraphics[scale=0.4]{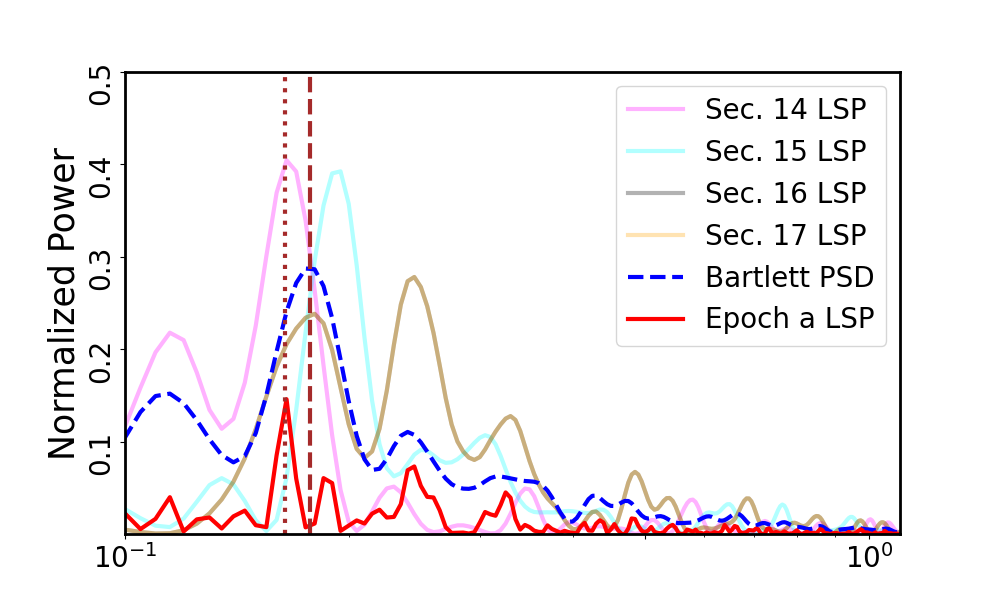}\includegraphics[scale=0.4]{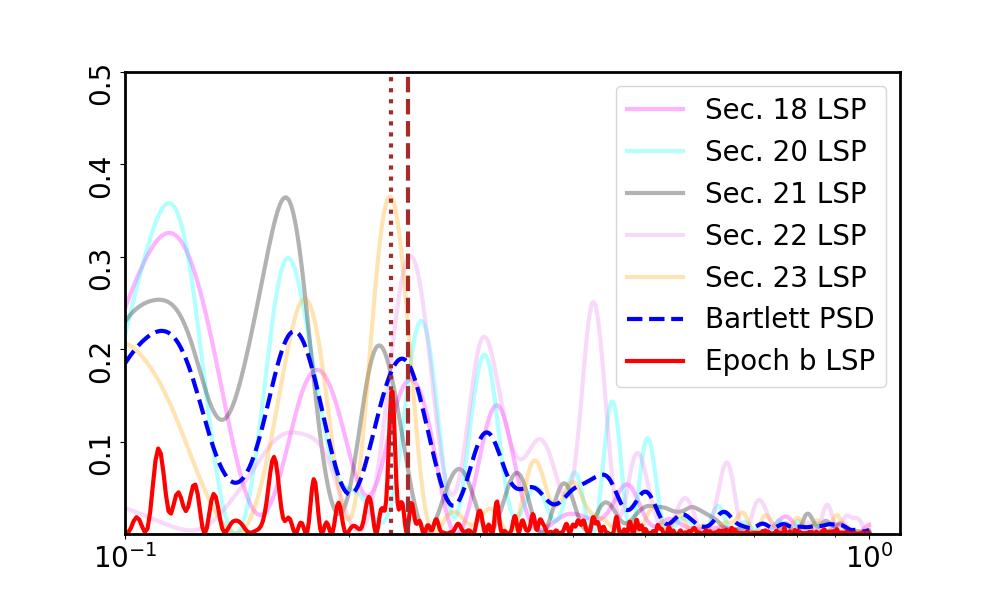} \includegraphics[scale=0.4]{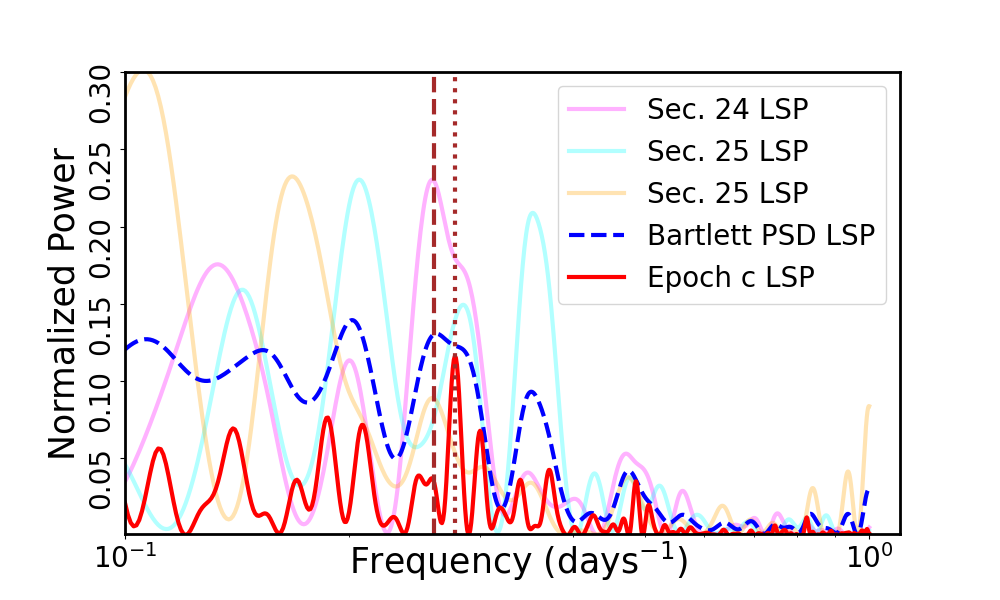}\includegraphics[scale=0.4]{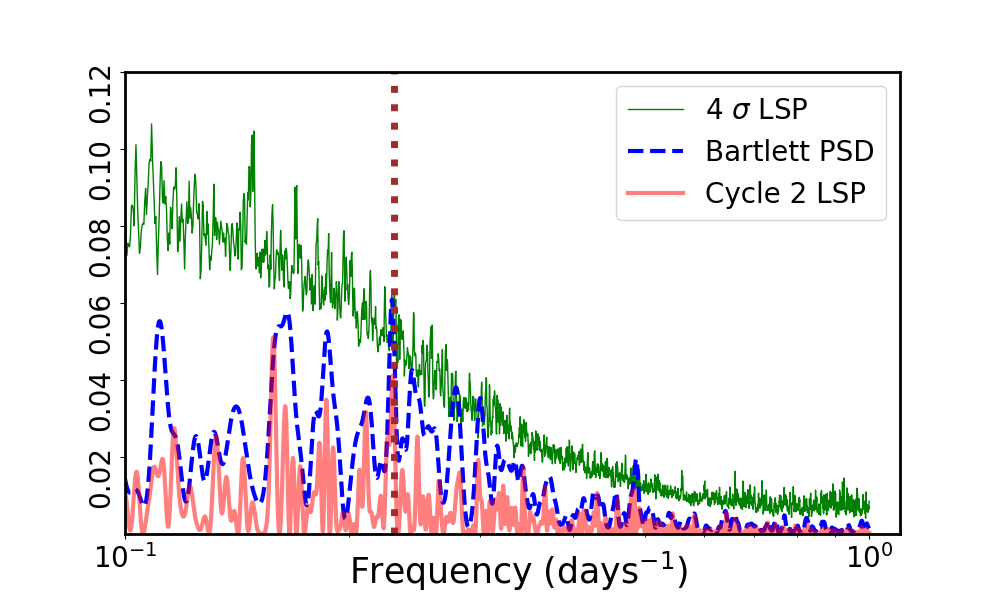}
\caption{ Barlett's periodogram (BP) for each epoch and the entire Cycle 2 observations. The Lomb-Scargle periodogram (LSP) for the epochs and corresponding sectors is also plotted for comparison. The vertical brown dotted curve shows the peak shown by LSP whereas the dashed brown curve shows the peak calculated by BP. For the full Cycle 2 observations, a 4$\sigma$ curve is also plotted, showing that the significance of the peak increases up to that level using the BP method, unlike for the simple LSP, presumably because of reduced fluctuations in the power spectrum. }
\end{figure*}\label{fig:bp}

\subsection{Mitigating gaps and de-noising: Bartlett method}

Figure~11 shows Bartlett's periodograms (BPs) estimated for each of the epochs of the TESS observations of 3C~371. We also show the PSDs of individual sectors. The BP of an epoch is calculated by averaging the periodograms of the included sectors, as described in Sect.~3.4. As it is a Fourier space analysis we can compute the BP for whole Cycle 2 observations by incorporating all 12 sectors. 

For epoch A, the peak signal of the average PSD is found at 0.18 d$^{-1}$, which is slightly different from that of the combined LSP (0.16 d$^{-1}$). The value of $\alpha_{high}$ is found to be $1.94 \pm 0.02$ which is a typical depiction of red-noise. The higher value obtained previously could be due to the variance or fluctuations that are smoothed out using this method. The strongest QPO signal is found to be at the frequency where the strongest signals from sectors 14 and 15 intersect, at 0.18 d$^{-1}$. Additionally, the significance of this peak improves and exceeds 4$\sigma$. The strength of the signals at lower frequencies increase as compared to those of the combined LSP. This method is particularly useful for epoch B where the long gap in the time domain could not be modeled properly by the C-ARMA process. In Fourier space, we can simply combine the periodograms using the BP method, without treating the Sector 19 gap. In the upper right panel of Fig.~11 we see that the BP yields the strongest peak at 0.24 d$^{-1}$, which is quite close to that found for epoch B by the LSP. The strength of the strongest signal improves marginally and the BP averages the noisy features, as in epoch A. For epoch C, the location of the strongest peak in the BP is at a lower frequency compared to what is found using LSP, as opposed to epochs A and B, where the BP's peak is a higher frequency than the LSP's. The QPO frequency for combined LSP is 0.28 d$^{-1}$, whereas for BP, the location is 0.26 d$^{-1}$. The strength of the signal increases marginally and remains above  4$\sigma$. So, in all three epochs, the BP approach reduces the variance of the periodogram and also increases the significance of the strongest signal along with other signals at lower frequencies. We also estimated the BP for the entire Cycle 2 observations by taking the average of the periodogram from all epochs.  Aside from reducing the variance as compared to the original LSP, the significance of a signal at $\approx$~6.3 days also improves to more than 4$\sigma$ using Bartlett's method. 

The main difference between BP and LSP is the greater strength of the peaks at lower frequencies found with the former. As the red-noise power is inversely proportional to a power of the frequency, the value of PSD power increases very rapidly at lower frequencies which could lead to the large amplification of noise beyond that of the true PSD \citep[see][and references therein]{2014MNRAS.445..437M, 2020ApJ...891..120B, 2024MNRAS.527.9132T}. So, the values of PSDs at lower frequencies should be taken with caution.

\begin{figure*}
\includegraphics[scale=0.4]{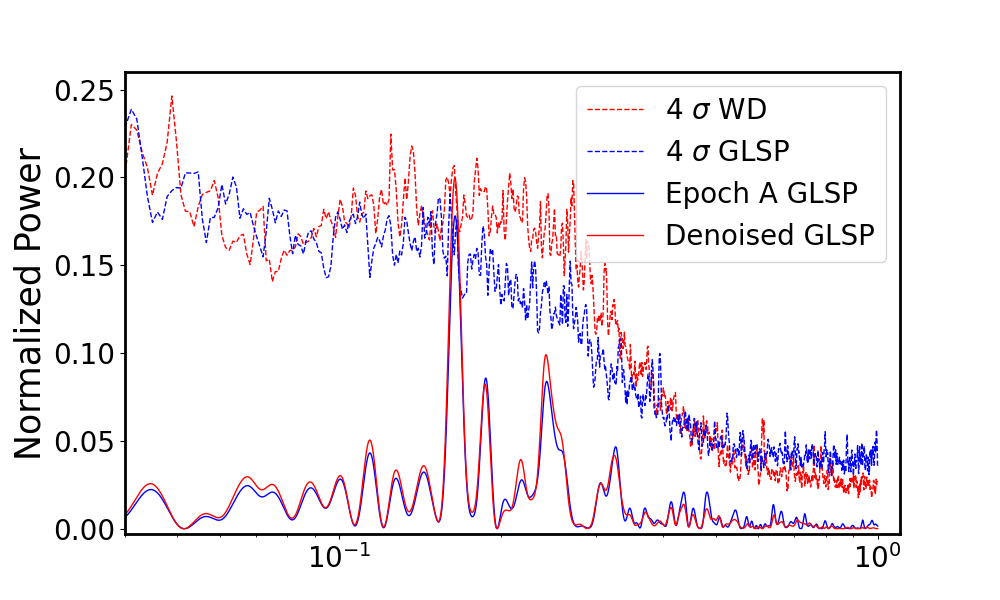}\includegraphics[scale=0.4]{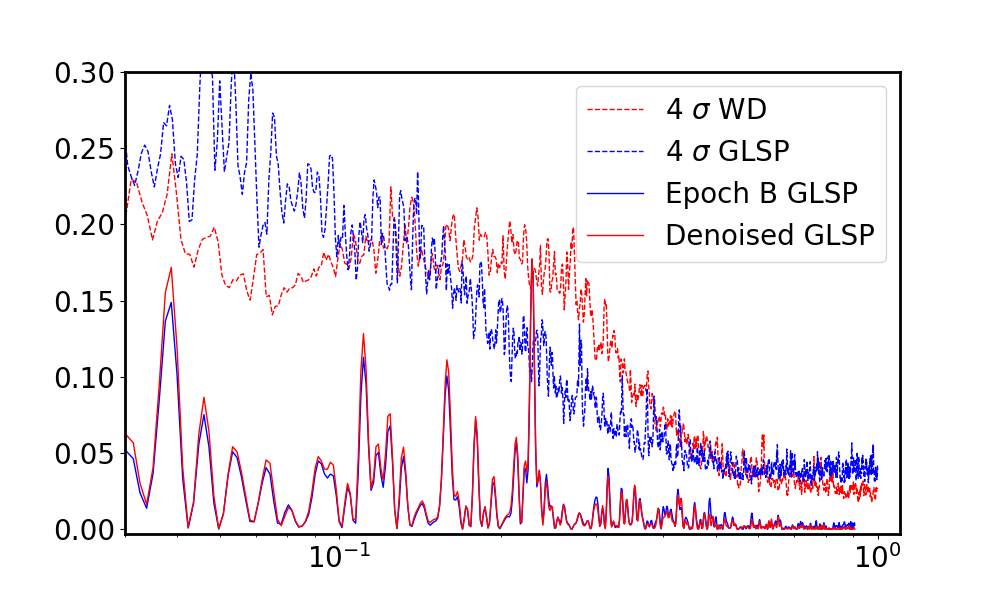}\\
\includegraphics[scale=0.4]{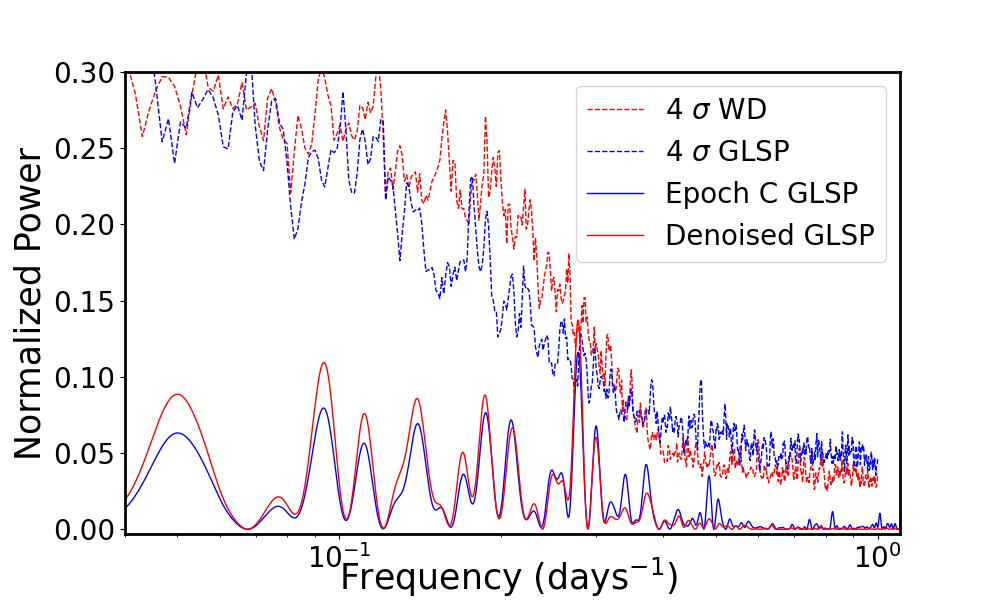}\includegraphics[scale=0.4]{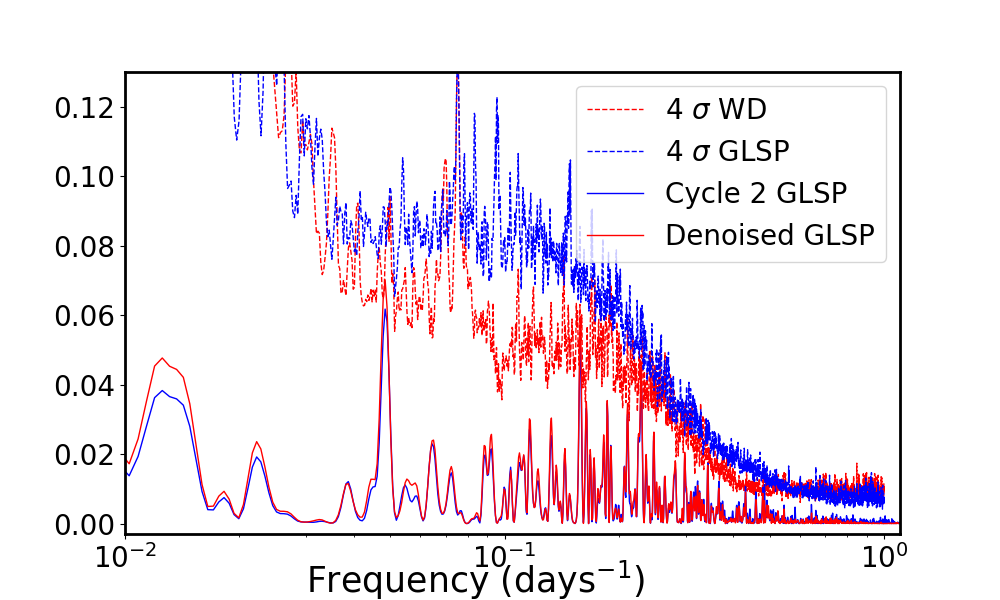}
\caption{The power spectrum (solid red curve) and the corresponding 4$\sigma$ significance curve (dotted red) were calculated using the wavelet decomposition (WD) method. The respective results from the GLSP method (in blue) are also plotted for comparison. The de-noised light curves for different epochs using WD are plotted in the appendix.}
\end{figure*}\label{fig:wd}

\subsection{De-noising the signal: Wavelet Decomposition (WD)}
De-noising the periodogram typically reduces the level of white noise in the PSD of the data. This improves the estimate of the red-noise levels and therefore better specifies the significance of a QPO signal, if present. In the WD method, one needs to fix the threshold corresponding to which the noise is reduced in the data  \citep[for details, see][]{2001C}. We employed a threshold such that it would reduce the variance/noise present, preserving the strength of the QPO signal in Fourier space. The threshold is very crucial in this method, as if it is too high, it will also reduce red-noise stochasticity along with white-noise. For these TESS observations, WD should prove helpful in detecting a real signal that is adversely affected by noise, especially at higher frequencies. 

The de-noised light curves using the wavelet decomposition method are plotted in Fig.~13. The minor fluctuations of the original light curves seem to be reduced, giving a smoother light curve that should be dominated by real stochastic variations. Figure~12 shows the PSD estimated using the GLSP method, which is obtained by analyzing the de-noised light curves generated by WD and the corresponding 4$\sigma$ significance curves for epochs A, B, C, and Cycle 2. For comparison, we also plot the GLSP for each epoch along with its 4$\sigma$ significance curve. For all epochs, the de-noised spectrum seems to have less power at higher frequencies as compared to that of the original PSD, indicating that the WD method proved effective in reducing the white noise. Besides, the strength of the peaks in the de-noised spectrum typically are stronger at lower frequencies ($f<0.2$ d$^{-1}$). As the claimed variability and periodicity timescales are around this frequency, this de-noising technique increases the strength of the peak corresponding to these timescales. At high frequencies, the WD curve seems to have a lower level of 4$\sigma$ significance  curve as compared to the original significance curve, thanks to the reduction in noise. As one goes towards lower frequencies, the WD curve becomes steeper and becomes higher than the original one. This is because decreasing power at higher frequencies and increasing power at lower frequencies have a compound effect on fitting the red noise with a broken power law, leading to a steeper spectral index. 
So, using the wavelet decomposition method effectively de-noises the power spectrum and consequently, the significance curve, which is substantially modified from the original GLSP result. At frequencies less than 0.1 d$^{-1}$, the difference in the original and WD significance curve is most pronounced in Epoch B, followed by Epochs C and A.  

The bottom right panel of Fig.~12 shows the results of the Cycle 2 observation. At a frequency higher than 0.5 d$^{-1}$, the levels of the WD and GLSP significance curves are similar. At lower frequencies, the WD significance level is lower than that of  GLSP, showing that this method reduces the noise present in the frequency range of 0.03--0.5 d$^{-1}$, leading to an improvement in the significance of the signals in the power spectrum. As this frequency range includes the variability timescales studied in this work, de-noising the light curve increases the significance of some signals. Peaks around 6 days are now found to be more than 4$\sigma$ significant. Despite using the same noise threshold, the frequency range for which the white noise is reduced for the entirety of Cycle 2 is different from the ranges in the case of individual epochs because the temporal baseline is 3 times longer for all of Cycle 2. 

\begin{table*}
 \caption{Variability timescales (in days$^{-1}$) obtained using different techniques used in this work for different epochs and whole Cycle 2 observation. }
\begin{tabular}{|c|c|c|c|c|c|}
\hline
Method & Epoch A& Epoch B & Epoch B1 & Epoch C & Cycle 2 \\ \hline
LSP  & 0.24 $\pm$ 0.02 &0.23 $\pm$ 0.02&0.23 $\pm$ 0.02 & 0.30 $\pm$ 0.04&0.23 $\pm$ 0.04\\
CARMA PSD &0.21 $\pm$ 0.02& 0.17 $\pm$ 0.03&0.14 $\pm$ 0.03& 0.24 $\pm$ 0.02&0.15 $\pm$ 0.03\\
SF (1/day) & 0.22 $\pm$ 0.02&0.24 $\pm$ 0.02&0.2 $\pm$ 0.03&0.23 $\pm$ 0.03&0.23 $\pm$ 0.03\\
CARMA SF (1/day) &0.2 $\pm$ 0.01&0.17 $\pm$ 0.01&0.19 $\pm$ 0.02&0.23 $\pm$ 0.02&-\\
Bartlett's PSD & 0.18 $\pm$ 0.03 & 0.24$\pm$0.01 & 0.24$\pm$0.03 & 0.28$\pm$0.01&0.24$\pm$0.06\\
WD&0.16 $\pm$ 0.02 &0.23$\pm$0.03&0.23$\pm$0.02&0.28$\pm$0.04&0.23$\pm$0.05\\ \hline 

\end{tabular}
\end{table*}

\section{Discussion}\label{sec:dis}

In this work, we have presented time-series analyses of TESS Cycle 2 observations of the blazar 3C 371 with 30 min cadence. The entire observation consists of 12 sectors, each of duration $\approx$~27 days, and so spans $\approx$300 days.  To produce the light curves we have used {\tt Quaver}, a program designed specifically for extracting TESS light curves of AGNs. The majority of sectors follow essentially Gaussian flux distributions. Sectors 16, 21, 23, 24, and 26 have tailed distributions, illustrating that these sectors have ``flare-like" processes affecting the variability. Despite having a high cadence and regular sampling of data, there exists an inevitable gap in the middle of every TESS sector observation because the telescope needs to turn towards the Earth to transmit its stored data;  the size of the gap varies from 1 to 5 days. 
Treating the effect of this gap is one of the most important aspects of TESS data analysis. 

We employed the C-ARMA process to fit the light curve directly in the time domain using stochastic differential equations. Solving these equations gives a prediction for the light curve during the gap. We used the C-ARMA (2,1) process, commonly known as a damped harmonic oscillator, and computed the corresponding second-order products such as power spectral density and structure-function. We calculated the variability timescales using C-ARMA products and compared them with the ones estimated using actual observations. The bending frequency, $\nu_b$, calculated by fitting the C-ARMA PSD with a broken power law, is also an indicator of the variability timescales present in the observation. It is found to be in the range of 0.13--0.32 d$^{-1}$ for individual sectors, which corresponds to 3--7 days' timescales. The first dip in the SF curve can also correspond to the variability timescale present in observation and is found to be in the range of 3--7 days, in most sectors, in agreement with those found using the PSD. It should be noted any over-fitting of the long-term behavior will lead to artificial flattening of the low-frequency end of the power spectrum. For TESS data, this effect becomes detectable around 7 days, which is one-fourth of the temporal baseline \citep{2023ApJ...958..188S}. The timescales obtained here are smaller than those obtained for most of the TESS sectors of 3C 371 analyzed in \citet{2020MNRAS.492.5524O}.  As discussed earlier, they calibrated their raw TESS data using ground-based observations. They obtained light curves similar to those produced by the simple hybrid light curve reduction option in {\tt Quaver}, which is preferable for examining variability on longer timescales, of the order of a sector's length.
Another effect of the suppression of long-term variability in the fully hybrid implementation are steeper $\alpha_{high}$ values. This is because the fully hybrid method removes more systematics which reduces the instrumental noise and consequently estimates a higher $\alpha_{high}$. 

To assess the effect of the gaps in the middle of TESS sectors on the variability timescales, \citet{2024MNRAS.527.9132T} simulated light curves with different lengths of gaps within the sectors and estimated the effect of these gaps on the distribution of PSD amplitudes of the periodograms. They calculated the simulated PSD distributions for when no gaps were present in the light curves and for when the gaps had lengths of 2--6 days. They found that the PSD distributions for all those cases agreed with each other for frequencies above 0.1 d$^{-1}$. Below 0.1 d$^{-1}$, the error estimates for the periodogram increase as the gap increases. 
As the variability timescales and the possible QPO periods are in the range of 3.0-7.0 days, which corresponds to the frequency of 0.14-0.25 d$^{-1}$, the gaps should produce minimal effects on the PSD distribution.

In Sectors 18 and 21, the C-ARMA SFs show no features, probably due to the low-order C-ARMA process we used. It is possible that higher-order C-ARMA processes could result in SFs with some features corresponding to the variability timescales. We also performed recurrence analysis. This RP method employs non-linear time series analysis to determine whether a particular state of a dynamical system recurs or repeats in the light curve, resulting in structured variability or periodicity. The RPs of almost all sectors show canonical checkerboard features distinctive of quasi-periodicity, which supports the presence of variability also seen in the PSD and SF analyses. Please note that the current recurrence analysis is qualitative. The next step would be to assess whether these RP features are quantitatively indicative of a stochastic or deterministic through an analysis of the line patterns in the RP (e.g., the lengths and distribution of lines throughout the RP as compared to canonical dynamical systems) and to assign significance values to those features \citep{2020MNRAS.497.3418P}. 

As many observations from continuous sectors are available for this source, it is important to explore whether the same type of variability is present when the sectors are combined  so that the source is at different overall flux levels. For this purpose, we divided Cycle 2 observations into three epochs, based on stationarity, so the whole Cycle 2 is divided such that sectors within each epoch have similar means.  The C-ARMA PSD gives the variability timescales to be in the range of 5.0-7.0 days for all three epochs and the whole Cycle 2,
which is in agreement with what we found for individual sectors. This timescale is in agreement with the SF results. There is a very large gap in Epoch B, as there is no data for Sec.\ 19. To assess the effect of such a big gap, we remove Sec.\ 18 and Sec.\ 19 to construct a new segment, called Epoch B1. The results for Epoch B and Epoch B1 do not differ significantly and display similar timescales. We also analyzed the whole Cycle 2 observation by making it a zero mean process, and the variability timescale found with C-ARMA analysis is 0.15 d$^{-1}$, which corresponds to 6.6 days. 


These variability timescales in the range of 3.0--7.0 days can be interpreted in a variety of ways.  They could correspond to the relaxation time related to synchrotron self-Compton emission, as described in \citet{2014ApJ...791...21F}.
Such variability timescales have also been reported in \citet{2019ApJ...885...12R} by studying the $\gamma$-ray emission in blazars, and it has been suggested that such timescales arise from instabilities inside the jet produced by a kink instability and magnetic reconnection \citep{2022Natur.609..265J}. Recently, \citet{2024ApJ...973...10D} also found similar timescales by analyzing TESS observations of southern blazars  The light curves analyzed in this work follow the red-noise stochasticity ($\alpha_{high} \approx 2$) which are consistent with models including leptonic emission and magneto-hydrodynamic simulations of turbulent jets \citep{2016ApJ...820...12P, 2019MNRAS.483..565C, 2019Galax...7...35T, 2024ApJ...973...10D}. In this work, both GLSP and WWZ analyses indicate the presence of multiple variability timescales, which could be the result of two physical processes acting in different emission regions. It is also possible that these processes may vary in intensity over time and alternatively dominate the observed light curve, resulting to shift in the characteristic variability timescales.
As these optical emissions of such small timescales are believed to originate in the jets, multiple timescales could arise from different processes related to plasma instabilities or the presence of sub-structure within a jet, or ``mini-jets" \citep[e.g.][]{2012A&A...548A.123B,2020ApJ...903..134C}.


Aside from the variability timescales, we also found the presence of a quasi-periodic signal whose significance is at least 99.99\% (4$\sigma$). 
QPO periods  of 6.0, 4.3, and 3.6 days are found for Epochs A, B, and C, respectively. These signals are found to persist throughout the respective epochs as is evident in WWZ results. The recurrence analysis also confirms the presence of variability on the QPO timescales. We used Bartlett's method as well as the C-ARMA method to mitigate the effect of the gaps. These methods work in different domains. C-ARMA fits the light curve in the time domain, whereas Bartlett's method averages the PSDs in Fourier space. Bartlett's method reduces the fluctuations or variance present in the observation and also improves the significance of the claimed QPO signal. For Cycle 2 analysis, a peak around 6 days is found to be more than 4$\sigma$ significant, while it was less significant in the simple PSD analysis. Interestingly, the bending frequency estimated from BP for all epochs and the whole Cycle 2 observations corresponds to the QPO frequency (as listed in Table 2), which could suggest the variability timescales arise from some fluctuation that got smoothed out by Bartlett's method. We also used the wavelet decomposition method, which 
reduces the white noise present in the higher frequency regime and improves the strength of peaks in the lower frequency regime, producing a more sophisticated treatment of noise, which also affects the significance level of the desired QPO signal. This method also improves the significance of a signal around 6 days for the full Cycle 2 observation. Like BP, the bending frequency also corresponds to the QPO frequency for all individual epochs for WD. 

The quasi-periodic signal, along with the variability timescale, suggests the presence of different kinds of physical processes or the same process having different timescales (mini-jets). The different QPO periods or variability timescales in different epochs can sensibly be taken to correspond to changes in the size of the emission region inside the jet  \citep{2012A&A...548A.123B, 2020ApJ...903..134C}. One of the most plausible models to explain the origin of such days' variability is the kink instability model \citep[e.g.][]{2022Natur.609..265J}. In this model, the variability is attributed to instabilities developed in the plasma of the jet, which has a toroidal magnetic field, as shown through magneto-hydrodynamical simulations \citep[e.g.][]{2020MNRAS.494.1817D}. The interaction between the instabilities and the toroidal magnetic field leads to magnetic reconnection and the formation of a kink at the center of a jet, which is quasi-periodic in nature. The temporal growth of the kink instability from the center of the jet to its boundary is manifested in the light curves as a variable quasi-periodic signature. \citet{2020MNRAS.494.1817D} estimated the growth time of this kink $T_{kink}$ as the ratio of the distance traversed by the kink from the center of the jet to its boundary, $R$, to the velocity of propagation, $\left< v_{tr} \right>$, and can be written as \citep{2020MNRAS.494.1817D, 2024MNRAS.527.9132T, 2024MNRAS.528.6608T}

\begin{equation}
T_{kink} = \frac{ R}{ \left< v_{tr} \right>\delta }~,
\end{equation}
where $\delta$ is the Doppler factor. For a typical blazar, the radius of the emission region in the co-moving frame is taken to be $10^{16}$--$10^{17}$ cm and the value of $\delta$ to be 10.0, following \citet{2024MNRAS.527.9132T, 2024MNRAS.528.6608T}. The value of $\left< v_{tr} \right>$ is probably of the order of 0.16$c$, as found in \citet{2020MNRAS.494.1817D}. Putting these values in the above equation estimates the growth time to be 1--10 days, which is consistent with the QPO and variability timescales found in this work. 

Apart from the kink instability and the presence of mini-jets, there are other processes in AGNs that might explain the quasi-periodicity of 3-7 days' timescales found in this work. The Lens-Thirring precession of the inner portion of the accretion disk around the SMBH could produce QPOs of the order of a week \citep{1998ApJ...492L..59S}. Another possible model is related to the normal modes of oscillations, which are trapped in the innermost region of the accretion disk because of the strong gravity produced by the compact object at its center\citep[e.g.,][]{1997ApJ...476..589P,2008ApJ...679..182E}. Magnetorotational instabilities (MRI) and the consequent turbulence could also produce periodicity of such timescales \citep[e.g.,][]{2004ApJ...609L..63A}. These optical QPOs might also be the result of X-ray reprocessing by the optical disk region \citep{2018ApJ...860L..10S}.

\section{Conclusions}\label{sec:con}

This paper has analyzed $\sim$300 days long TESS observations of the blazar 3C 371. This nearly continuous observation has been proven very effective in detecting the presence of variability of around 6 days which can be associated with the kink-instability or mini-jets. We also found the variability timescales in the range of 3.0-7.0 days in 12 sectors constituting the Cycle 2 observation. Interestingly, we found the signature of probable QPOs in different epochs of this observation. In this work, besides studying the variability and quasi-periodicity, we presented methods that can be used to handle issues (gaps, noise)  in TESS data, such as C-ARMA modeling, Bartlett's method, and wavelet decomposing. These methods estimate the model parameters that are in agreement with the values found in the literature for stochastic AGN variability.  The different timescales for variability and periodicities in different epochs could signal the presence of different underlying physical processes.

TESS has proved exceptionally useful for the study of AGNs as well as for its primary purpose of exoplanet detection. The high-cadence, regularly sampled data make it an excellent tool for probing short-term variations in many AGNs. And for a small number of sources, such as 3C 371, where TESS observations span many consecutive sectors, it allows probing longer-term variations as well. To take advantage of such high-quality data, the inevitable gaps present within the sectors or between them and the noise affecting the AGN variability should be treated carefully.  Our study concludes that the methods presented in this work are indeed effective in countering these issues and could certainly be used to analyze multiple-sector TESS observations of other blazars as well as Seyfert galaxies in future work.   \\

We gratefully acknowledge the referee for remarks that significantly improved the manuscript. This work was supported by the Tianshan Talent Training Program (grant No. 2023TSYCCX0099), the CAS `Light of West China' Program (grant No. 2021-XBQNXZ-005), and the National SKA Program of China (grant No. 2022SKA0120102). A.T. acknowledges the support from the Xinjiang Tianchi Talent Program. This work was also supported by the Urumqi Nanshan Astronomy and Deep Space Exploration Observation and Research Station of Xinjiang (XJYWZ2303).


 \section*{Data Availability}
The TESS data presented in this paper were obtained from the Mikulski Archive for Space Telescopes (MAST) at the Space Telescope Science Institute.

{}

\appendix

\section{Wavelet decomposition}
Fig.~13 shows the de-noised light curve in red for epochs A, B, and C. The original light curve is also plotted in black for comparison. Using WD essentially reduces the high-frequency noise in the periodogram and also hints at the presence of possible variability or quasi-periodicity in the data. As WD reduces the noise, it could also allows for the calculation of the significance relative to the red noise stochasticity in a more sophisticated manner as compared to the original, noisy light curve. 


\begin{figure*}
\includegraphics[scale=0.4]{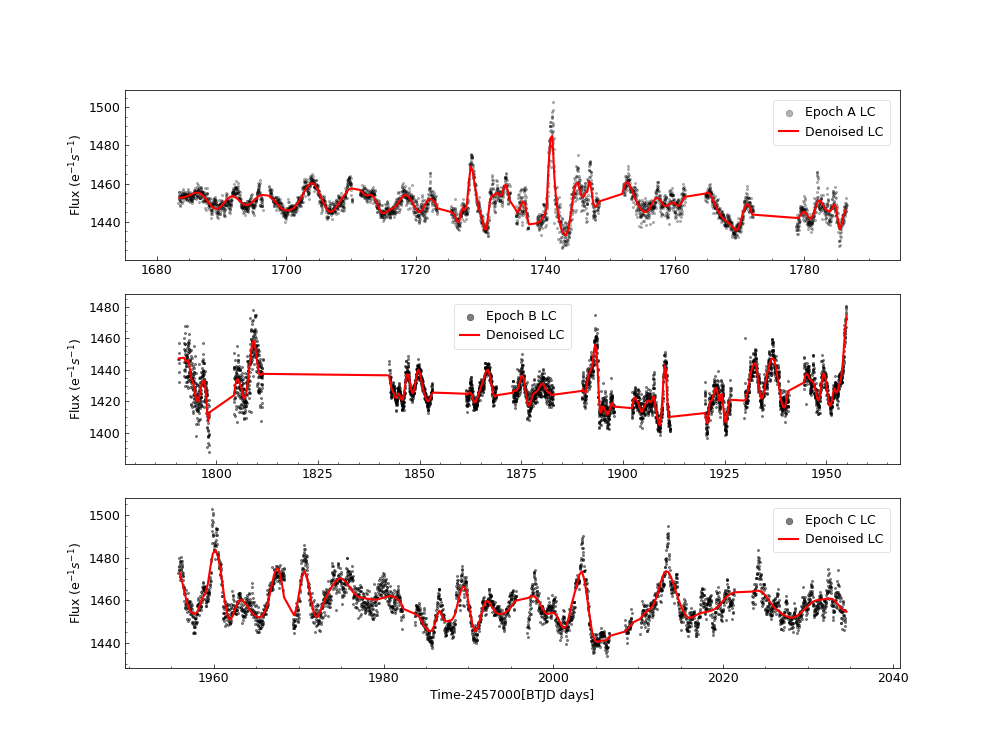}
\caption{ De-noised light curves (in red) produced using wavelet decomposition for epochs A, B, and C. The original light curves (gray points) are also plotted for comparison.} 
\end{figure*}\label{fig:wdlc}

\end{document}